\documentclass[twocolumn,showpacs,floatfix,prb,aps,10pt,superscriptaddress]{revtex4-1}
\usepackage{epstopdf}
\usepackage{graphicx}
\usepackage{color}
\usepackage{enumerate}
\usepackage{amsmath}
\usepackage{amssymb}
\usepackage{bbm}
\usepackage{amsfonts}
\usepackage{subfigure}
\usepackage{bm}
\usepackage{multirow}
\usepackage{bbm}
\usepackage{comment}
\usepackage{longtable,booktabs}
\usepackage{hyperref}
\usepackage{url}
\usepackage{subfigure}
\usepackage{dsfont}
\newcommand{\m}{\mathbf{m}}
\newcommand{\bb}[1]{\mathbf{#1}}
\usepackage{float}

\newcommand{\ket}[1]{\ensuremath{\left|#1\right\rangle}}
\begin{document}
\title{Fermion Parity Flips and Majorana Bound States at twist defects in
Superconducting Fractional Topological Phases}
\author{Mayukh Nilay Khan}
\affiliation{Department of Physics and Institute of Condensed Matter Theory, University of Illinois at Urbana-Champaign, IL 61801, USA}
\author{Jeffrey C. Y. Teo}
\affiliation{Department of Physics, University of Virginia, Charlottesville, VA 22904 USA}
\author{Taylor L. Hughes}
\affiliation{Department of Physics and Institute of Condensed Matter Theory, University of Illinois at Urbana-Champaign, IL 61801, USA}
\author{Smitha Vishveshwara}
\affiliation{Department of Physics and Institute of Condensed Matter Theory, University of Illinois at Urbana-Champaign, IL 61801, USA}
\begin{abstract}
In this paper we consider a layered heterostructure of an Abelian topologically ordered
state (TO), such as a fractional Chern insulator/quantum Hall state with an s-wave
superconductor in order to explore the existence of non-Abelian defects. In order to uncover such defects we must augment the original TO by a $\mathbb{Z}_2$ gauge theory sector coming from the s-wave SC. We first determine the
extended TO for a wide variety of fractional quantum Hall or fractional Chern insulator heterostructures. We prove the existence of a general anyon permutation symmetry (AS) that exists in any fermionic Abelian TO state in contact with an s-wave superconductor. Physically this permutation corresponds to
adding a fermion to an odd flux vortices (in units of $h/2e$) as they travel around the associated topological (twist) defect. As such, we call it a fermion parity
flip AS. We consider twist defects which mutate anyons according to the fermion
parity flip symmetry and show that they can be realized at domain walls between distinct
gapped edges or interfaces of the TO superconducting state. We analyze the properties of such
defects and show that
fermion parity flip twist defects are always associated with Majorana zero modes. Our formalism also
reproduces known results such as Majorana/parafermionic bound states at
superconducting domain walls of
topological/Fractional Chern insulators when twist defects are constructed based on charge conjugation symmetry.
Finally, we briefly describe more exotic twist liquid phases 
obtained by gauging the AS where the twist defects become deconfined anyonic excitations.
\end{abstract}
 \maketitle 

\section{Introduction}
Recent years have seen an explosion of interest in solid-state systems 
having excitations that obey non-Abelian statistics. These non-Abelian
excitations are sought-after because of their potential for decoherence free
topological quantum
computation (TQC)
\cite{Kitaev03,OgburnPreskill99,Preskilllecturenotes,FreedmanKitaevLarsenWang01,Wangbook,
ChetanSimonSternFreedmanDasSarma,SternLindner13}. Early theoretical proposals
of non-Abelian zero excitations include quasi-particles in the 5/2 Pfaffian quantum Hall state,
\cite{MooreRead,ReadGreen,GreiterWenWilczek91,SternOppenMariani,NayakWilczek96}
half quantum vortices in $p+ip$ superconductors 
\cite{Ivanov,StoneChung}, and the boundary modes
modes of a one-dimensional $p$-wave superconducting wire \cite{Kitaevchain}.

While the verdict is still out on the presence of these excitations in the aforementioned systems, one focus of the field has moved to the study of semiclassical defects
having non-Abelian statistics
in engineered superconducting
heterostructures.
Beginning with the seminal work of Fu and Kane \cite{FuKane08}, there has been
an intense interest in understanding the properties of layered heterostructures
of  topological (and some non-topological) phases and proximity-coupled superconductors. 
In fact, one experimental
realization of  an effective 1D Kitaev $p$-wave wire has been proposed, and experimentally sought-after, in spin-orbit coupled materials in superconducting 
proximity to an s-wave superconductor.\cite{SauLutchynTewariDasSarma,OregRefaelvonOppen10,
Kouwenhoven12,Sato2009,DengYuHuangLarssonCaroffXu12,Shtrikman12,RokhinsonLiuFurdyna12,
ChangManucharyanJespersenNygardMarcus12,FinckHarlingenMohseniJungLi13,NadjDrozdovLiChenJeonSeoMacDonaldBernevigYazdani14}.
Additionally, in higher dimensions, one of the seminal
and promising proposals  is the ability to realize non-Abelian
Majorana bound states (MBS) localized on vortex defects in a superconductor
placed on the surface of a 3D time-reversal invariant topological insulator
(TI)\cite{FuKane08}.  In general, the ability to combine two simpler phases without non-Abelian excitations, and as a
result allow for defects with non-Abelian statistics, is an important discovery
for the field of TQC. 

Along the lines of the focus of our article, Fu and Kane also proposed that MBS could 
appear in a 2D quantum spin Hall insulator (QSH) at an edge where a domain wall between superconductor and
magnetic coatings is formed\cite{FuKaneJosephsoncurrent09}. Knowing this, it was natural to generalize
these results to the study of proximity effects in other 2D interacting topological
phases.  Remarkably, several groups have predicted that similar domain walls in
 fractional quantum Hall (FQHE)/Chern Insulator (FCI) states can trap generalized MBS (or
parafermions)\cite{Fendley12,fradkin1980disorder} with a quantum dimension $d=\sqrt{2/\nu}$ that depends on the
underlying topological
phase\cite{ClarkeAliceaKirill,LindnerBergRefaelStern,MChen,Vaezi}. As an
example, if one takes a $\nu=1/3$ Laughlin state as a starting point, then such
 domain-wall defects trap $Z_6$ parafermions with $d=\sqrt{6}.$ 
 
 The main results of our article rely on our generalization of these results to large classes of Abelian FQHE states of fermions. In particular, we show that for all such FQHE states, proximity-induced non-Abelian zero modes (NAZM) could give rise to just an ordinary MBS instead of a more exotic parafermionic mode. 
 The outcome of which NAZM is stabilized depends on the details of the interactions near the defect. 
Hence, even though one may optimistically hope for
 parafermions, non-universal interactions might drive a MBS to be stabilized. This prediction is a consequence of a previously-ignored anyonic symmetry transformation that exists in \emph{every} Abelian topological phase with local fermions.
 In addition to this experimentally motivated prediction, we also develop a theory that describes the interplay between fermionic FQHE states and superconductivity (through its connection with fermion parity symmetry) at an increasingly exotic level, which will be discussed further below.

\subsection{Motivation for Our Approach}
In parallel with current developments on proximity-induced defects, a rich
literature has developed on the subject of general types of semiclassical defects in topological phases \cite{Kitaev06,Bombin,
Bombin11, KitaevKong12, kong2012A,YouWen, YouJianWen, BarkeshliQi,
BarkeshliQi13, BarkeshliJianQi, MesarosKimRan13, TeoRoyXiao13long,
teo2013braiding,khan2014,Levin13,teodefectreview15,barkeshli2010,YouJianWen,MesarosKimRan13,Petrova14}, some examples being dislocations and disclinations in lattice
or bilayer/trilayer systems. In each case, the properties of the defects in question are determined by the
nature of the topological properties of the bulk. However, often the structure of the defects is uncovered from the edge properties since a defect can be represented as a mass domain wall on a corresponding gapped edge or a seam/interface that has been glued back together. A defect forms when the edge or seam is gapped out in two distinct ways, hence leading to a domain wall
with a possible NAZM. 
For example, on the edge of the QSH insulator there exists a free, massless
Dirac fermion, and this can be gapped out via magnetic or superconductor proximity coupling. An interface between the two gaps is exactly what was mentioned above, and binds a MBS. Other topologically ordered phases will have more general edge conformal field theories (CFTs), and the types of defects can be classified by the allowed gapping terms of CFTs ( if they form a non-chiral edge), or these CFTs combined with a time-reversed partner (say at a seam or interface).

Some types of defects act to permute orbiting anyons that exist in a parent topologically ordered state. We will call these \emph{twist
defects}, and the associated discrete permutation symmetry of the anyonic excitations of the 
underlying topologically ordered (TO) phase we will call an anyonic symmetry (AS). Semiclassical twist defects are attached to branch cuts across which anyon labels are acted upon by the action
of the AS group. These permutations exchange different anyons, but leave
their fusion and statistical properties
invariant. Interestingly, due to their unusual nature, a twist defect traps a NAZM at its core, and indeed the MBS discussed above can be thought as sitting at the core of a twist defect.

Since it is usually the NAZM at the defect core that is of interest for TQC, it
is important to be able to identify their fusion properties, e.g., their quantum dimension. Recently, a wide variety of general
techniques has been developed to understand the nature of NAZM at twist defects, 
\emph{without} resorting to arguments based on the 
edge CFT. \cite{TeoRoyXiao13long,BarkeshliJianQi,khan2014}.
In particular, a clear algorithm was developed: 
\begin{align*}
 \text{TO media}\xrightarrow{\text{AS group}}\text{Twist
 defects}\xrightarrow{\text{Fusion category}}\\
 \text{ Quantum dimension of NAZM at defect.}
\end{align*}\noindent In this diagram we begin with some topologically ordered phase and then identify the possible AS operations that permute anyons. To each non-trivial permutation element there corresponds (at least) one twist defect. If the fusion structure of fusing a twist defect with other defects (including ones similar to itself) can be calculated, then the details of
the NAZM core can be extracted. For example, if a twist defect fuses with a defect of the same type to give two anyons, each with quantum dimension 1, then a single twist defect will carry a NAZM with quantum dimension $\sqrt{2}$, i.e., a MBS. 

Gauged versions of these
semiclassical defects, where the defects are deconfined and become fully
evolved quasiparticles were also
considered\cite{barkeshli2010,barkeshli2014,Teotwistliquids,tarantino2015symmetry}.
When deconfined, the twist defects also generate a  modular braiding structure, not just a fusion structure as befits semiclassical defects, thus enriching the original topological phase by becoming genuine (non-Abelian) quantum excitations of a new phase with enhanced topological order. 

Before moving on to the summary of results, let us make some comments about our
approach in order to avoid possible confusion. The goal of our article is to
predict phenomena in FQHE/Superconductor heterostructures. It is known that
parafermionic NAZMs can exist in these systems. Interestingly, we will show that if the
FQHE has fermionic local quasiparticles then it can also host just conventional MBS
as well. Unfortunately, since FQHE states are strongly interacting systems,
determining the nature of defect bound states is challenging, unlike the case
of a non-interacting Chern insulator which can be handled using single-particle
exact diagonalization. Hence, in order to find the properties of
superconducting proximity-induced bound states, we will have to use a bit of an
indirect method. Our method essentially involves treating the superconductor
itself as a system with deconfined anyons\cite{HanssonOganesyanSondhi04} ($\mathbb{Z}_2$
topological order analogous to the toric code) and considering the interplay
between those anyons ($\mathbb{Z}_2$ charges and fluxes) and the proximity coupled FQHE
state. From here we will show that this combination of both the $\mathbb{Z}_2$ and the
FQHE anyons has a new AS connected to fermion parity, which directly
corresponds to a twist defect having a MBS. We also show that the general existence of a charge conjugation anyonic symmetry predicts the existence of the usual parafermionic twist defects. In the end, our predictions only rely on features available in ordinary superconductors when coupled to FQHE states. This is somewhat in the same spirit as Refs. \onlinecite{LevinGubraiding,Braiding3DLevinWang,LevinWang3Dbraiding} where the authors gauge a global symmetry to glean additional information about the un-gauged, symmetry-protected theory. Here we end up gauging a symmetry to learn about hidden twist defect structure that still survives in the un-gauged theory.
 \begin{align*}
 \text{Fermionic FQH}\xrightarrow{\text{gauge $\mathbb{Z}_2$}}\text{Extended bosonic TO}\\\xrightarrow{\text{AS}}\text{Twist defects}
\end{align*}

A side-effect of this approach is that on the way to our final goal we will derive properties of some more exotic topologically ordered states, which are of interest, but are not immediately experimentally relevant. These results include the study of gauging the $\mathbb{Z}_2$ fermion-parity symmetry of a wide variety of Abelian FQHE states, identifying the resulting AS operations of these gauged theories, and then working out the anyon content of the non-Abelian topologically ordered phase resulting from gauging the AS (deconfining the twist defects).


\subsection{Summary of main results} While there is a clear way to understand
NAZM in TO media using the technology of twist defects and
anyonic symmetries, a well defined theory for twist defects in the
aforementioned superconducting TO heterostructures is yet to be formulated. The excitations of the superconductor (vortices, quasi-particles) are not deconfined excitations, and thus the premise on which the usual theory of twist defects is based does not apply. Showing how one can overcome this problem is  one of the main objectives of this paper. While there has been some progress to this end
in Ref. \onlinecite{BarkeshliJianQi}, we provide a more thorough way to treat the problem by considering the full
AS in a system with topological order augmented by a $\mathbb{Z}_2$ gauge theory sector coming from the superconductor. In fact, ignoring this kind of extended TO due to the presence of
the s-wave superconductor
obscures a particular AS group operation, the fermion parity flip, which is a generic anyonic symmetry in fermionic systems, similar to charge conjugation. In short, we provide a unified approach to understanding 
twist defects in layered superconductor/TO heterostructures. 

In particular, we focus on layered heterostructures of a superconductor and an Abelian TO
state with local fermion excitations, e.g., the $\nu=1/3$ Laughlin state formed in a 2D electron gas or fractional Chern insulator. The superconductor can harbor confined defects/excitations including vortices, Bogoliubov de-Gennes (BdG) quasiparticles, and their composites. We determine the fusion structure of the anyon quasiparticles (qps) with these confined objects, 
and show that it is identical to that of the original TO phase, but with its $\mathbb{Z}_2$
fermion parity symmetry gauged (i.e., equivalent to the phase with deconfined  $h/2e$ fluxes of the superconductor).
Crucially, we note that the nature of twist defects in such heterostructures depends
only on the fusion structure (i.e., not braiding) of the qps and confined objects. Thus
even though we will often use the fermion parity gauged
theory for our description of the twist defects, the resulting properties hold true in the original heterostructure with an ordinary (``un-gauged") superconductor. Throughout our article we develop the theory of these fermion-parity gauged systems and their extended TO.
We give a field theoretic description of this procedure and several
examples relevant to spin singlet states, the Jain series at observed filling
fractions, and quantum Hall hierarchy states derived from $ADE$ Lie algebras\cite{Readhierarchy,FrohlichZee,
FrohlichThiran94,FrohlichStuderThiran97,cappelli1995stable}. 

Given this theory with $\mathbb{Z}_2$-extended TO we can then apply the usual twist defect formalism to this new theory.  As mentioned, one of our main results is that 
we find that \emph{any} fermionic Abelian TO state with a superconductor proximity effect has a fermion
parity flip anyon relabeling symmetry. To our knowledge this symmetry has been ignored to date.
One manifestation of this symmetry is in terms of a new class of
twist defects which can host MBS. We place this result in the context of
aforementioned previous work
\cite{ClarkeAliceaKirill,LindnerBergRefaelStern,MChen,Vaezi} where defects
between superconducting and normal media in Laughlin $\nu=1/m$ states harbor
$\mathbb{Z}_{2m}$ parafermions. We show that the parafermionic twist defects can also be generally
understood in our formalism, but using another AS: charge conjugation. 
Importantly,
our results imply that non-universal interaction terms near the defects can favor MBS
over parafermions in the same heterostructure.  In other words, they may favor
the fermion parity flip symmetry over charge conjugation.  We explicitly
construct examples of such interaction terms in the case of the $\nu=1/3$ state.  We
note that our results also hold for Chern insulators, and even non-chiral TIs, even though they
are not TO. In particular Ref. \onlinecite{teo2015topologically} previously considered the fermion
parity flip symmetry for the CI and QSH cases in detail. In some respects this paper may be 
considered as a generalization of those results to interacting systems.

The AS groups referred to above are global (but unconventional\cite{lu2013classification}) symmetries of the system. The twist defects act as fluxes
of an AS group which act on the topological charge labels of the anyons (via an AS group element), thus permuting them. One can gauge the AS,
thus deconfining the twist defects and leading to an exotic non-Abelian state called a twist liquid.
To complete our discussion, we use
the procedure outlined in Refs. \onlinecite{MesarosKimRan13,barkeshli2014,Teotwistliquids} to
determine the structure of two resulting families of twist liquids derived from gauging either the fermion parity flip
or charge conjugation symmetries in any one-component Laughlin state.

Our paper is organized as follows. In section \ref{sec:LaughlinFusionCategory} we
illustrate our main results for systems with the same TO as one-component fermionic Laughlin states. We
determine the anyonic fusion structure that results from the presence of a proximity-coupled
superconductor, paying close attention to the fact that the $h/2e$ flux vortex
is not deconfined in real superconductors. We then determine the AS/twist defects of the heterostructure are
effectively the same as that of the original theory but with gauged fermion parity symmetry, where the $h/2e$
vortex is deconfined. Focusing on the fermion parity flip
AS, we end the section with a discussion of the corresponding twist defects in these
Laughlin states. 

In Section \ref{sec:gaugingfermionparity} we construct a gauge theory
where we couple the Chern Simons action of a TO state with a $\mathbb{Z}_2$ gauge
theory of a dynamical superconductor and we determine the properties of the resulting field theory.
We enumerate some important physical properties and consistency checks that this theory satisfies, and show the
calculation specifically for the Laughlin states; we give many other examples in Appendix \ref{sec:examples}. In Section
\ref{sec:fermionparitygeneral} we show the existence of a fermion parity flip anyonic symmetry for any
Abelian TO state having local fermions. In Section \ref{sec:AS} we consider a quasi-one dimensional edge/interface of a
TO state and show how this AS can be actually realized at the gapped edge/interface via
backscattering terms. Section \ref{sec:twistdefect} introduces 
twist defects more systematically and computes the non-Abelian quantum dimension
and multichannel fusion rules for the fermion parity flip twist defect. Finally,
we end with  Section \ref{sec:Asgauge} which describes how to gauge the AS associated with the
twist defects. There are several appendices which explicitly show how to
determine the fermion parity gauged theory for a wide variety of fractional
quantum Hall hierarchy states, discuss issues of relevance of gapping terms 
at the quasi-one dimensional edge/interface, and provide alternative perspectives on the
emergence of the fermion parity gauged theory.

\section{Laughlin State in Proximity to an s-wave Superconductor  }\label{sec:LaughlinFusionCategory}
In this Section we provide an illustration of the methods and results of our article using the simplest set of TO states, the one-component Laughlin states. We will carry through the full program for these states (except for gauging the AS which we do in Sec. \ref{sec:Asgauge}) as a helpful example to keep in mind when reading through the more technical sections afterward.

Let us begin with the effective theory of either a FQHE 2DEG or fractional Chern insulator (FCI) (with the same TO as the Laughlin
FQH state) at filling $\nu=\frac{1}{2n+1}$.  It is described by the $K$-matrix
$K=2n+1,$ and a charge vector $\mathbf{t}=1$ \cite{WenZee92} (for a review of the
$K$-matrix formalism see Appendix \ref{app:Kmatrix}). Laughlin's argument
dictates that the fundamental quasiparticle excitation, which we denote by $\mathcal{E},$ is bound to an
$h/e$ flux, and hence carries an electric charge of $e^\ast=\nu $ (in units of $e$)
\cite{Laughlinargument,Laughlin83}. Its spin-exchange statistics are given by
$h_{\mathcal{E}}=\nu/2,$ or equivalently a statistical angle of
$\delta_{\mathcal{E}}=2\pi h_{\mathcal{E}}=\pi\nu$ \cite{ArovasWilcezkSchrieffer}, which in this case can be determined purely from Aharonov-Bohm physics.  An electron can be written as the composite of $2n+1$ $\mathcal{E}$ qps:
$\psi$ = $\mathcal{E}^{2n+1}$.  Additionally, since we are considering a TO state built from local fermions,  all
qps braid trivially around the electron.

Now let us add another component to our system by considering the Laughlin
state placed in proximity to an $s$-wave superconductor ($s$-SC), e.g., by
depositing a superconductor on a 2DEG or FCI material. Since, for simplicity,
we are only considering an s-wave superconductor one might worry about the
effect of spin polarization on the strength of a proximity effect. To combat this we must
either consider something like engineered domains with oppositely signed g-factors, as was recently
exploited in Refs. \onlinecite{ClarkeAliceaKirill} to allow for a proximity effect in a
2DEG Laughlin state, or consider an FCI where the Chern band degrees of freedom
are spin.  In both cases an interface, or a pair of
counter-propagating edge states, can be designed such that it will contain
the necessary degrees of freedom to support an $s$-wave proximity effect. For simplicity we will usually consider the FCI case as shown in Fig. \ref{fig:SCTOlayer}. 

Heuristically we can imagine that the presence of the superconductor leads to an effective
additional fractionalization as there are ``smaller" fundamental excitations given by the $h/2e$
vortex (henceforth referred to as $\m$), which carry a charge of only
$q_{\m}=e^\ast/2=\nu /2$ (modulo $2$).  The vortex $\m$  braids around the electron $\psi$ with a
phase of $-1$, thus an electron is no longer a local particle and can be \emph{topologically} distinguished from the trivial vacuum by its braiding with $\m.$  Thus the vacuum consists of Cooper pairs of charge $2e$, which form the new local, and bosonic, particles. With this in mind,  the full qp content of the combined systems
  (modulo local Cooper pairs) is generated by $\m$ and given by $1 (\equiv
\m^{8n+4}),{\bf m},{\bf m}^2\equiv \mathcal{E},\m^3, \ldots,{\bf m}^{8n+3}.$  The Laughlin qp $\mathcal{E}\equiv{\bf m}^2$ and the
electron $\psi\equiv{\bf m}^{4n+2}$.

 We now have a theory with a well-defined qp fusion group $\mathbb{Z}_{8n+4}$;
 $\m^{a}\times\m^{b}=\m^{a+b\text{ mod }(8n+4)}$, but we have gotten a bit ahead of ourselves.  In a real SC the $h/2e$ vortices
 are not strictly deconfined excitations and actually interact with each other, and
 thus do not have well defined exchange statistics.  Hence, they cannot serve as conventional anyons. However, the subset of qps generated by the $h/e$ flux ($\m^2$)
 $\mathcal{E}^{a}\equiv{\m}^{2a},a\in \mathbb{Z}$  of the original Laughlin state are still deconfined with well
 defined exchange and braiding. Additionally, fusion between the original qps and the vortices in the $s$-SC is also well-defined and, in fancier terms, we have a partially braided fusion category. At this stage we could consider gauging the $\mathbb{Z}_2$ fermion parity symmetry that would allow for a deconfined $\m$ qp, along with the other anyons of the $\mathbb{Z}_2$ topological order. In this case we would have a full braided fusion category and a new TO order would arise described by the K-matrix $K_{\mbox{sc}}=8n+4.$ Let us hold off on doing this for now and we will discuss it in more detail later. 


 Now, we search for permutations $\mathcal{P}$ of the full qp set
 $\mathcal{A}=\{1 (\equiv \m^{8n+4}),{\bf m},{\bf m}^2,\ldots,{\bf m}^{8n+3}\}$
 such that the map $\mathcal{A}\rightarrow \mathcal{PA}$ preserves the fusion
 algebra and the braiding statistics (of the deconfined particles). As $\mathcal{P}$ 
 preserves fusion, and $\mathcal{A}$ is generated by $\m$ alone, then $\mathcal{P}$ can be completely specified by its action on $\m$.
Hence, without loss of generality, let $\mathcal{P}\m\equiv \m^p;p\in\mathbb{Z}^{+}$. Further, bijectivity of $ \mathcal{P}$ (permutations are always bijective) requires
 that $p$ and $8n+4$ are mutually prime, i.e., gcd($p,8n+4$)=1. Thus, $p$ is
 odd. 
 
Now, let us consider the constraints imposed in order to have
 invariance of the braiding statistics. We have
 $\delta_{\mathcal{E}}=\frac{\pi}{2n+1},$ and under $\mathcal{P}$ we find
 $\delta_{\mathcal{P}\mathcal{E}}=\frac{p^2\pi}{2n+1}$. We want to impose the constraint: \begin{align}
\delta_{\mathcal{E}}=\delta_{\mathcal{P}\mathcal{E}} \text{ (mod)
}2\pi\nonumber\\ \implies p^2=1 \text{ (mod) } 4n+2\nonumber\\ \text{$p$ is odd
}\implies p^2=1\text{(mod) } 16n+8 \label{eqn:eqnnoname1} \end{align}\noindent where the last line is the simplified form of the constraint.

We will give some examples of $p$ satisfying this constraint below, but let us make some important comments. Suppose that we were to treat $\m$ as a genuine deconfined qp
with well defined exchange statistics. Then using $\mathcal{E}=\m^2$, the statistical phase of $\m$ is
$\delta_{\m}=\pi\nu/4=\frac{\pi}{8n+4}\implies h_{\m}=\pi\nu/8$.  This
is consistent with the Abelian topological state $K_{\mbox{sc}}=8n+4$ mentioned above, with charge vector
 $\mathbf{t}_{\mbox{sc}}=2$ (charge of a Cooper pair), and with the same
 fusion group $\mathbb{Z}_{8n+4}$ of our partially braided fusion theory.
 Importantly, we see that the constraint \eqref{eqn:eqnnoname1} automatically ensures
 $\delta_{\m}=\delta_{\mathcal{P}\m}$.  Thus, $\m\rightarrow \m^p$ with $p$
 obeying \eqref{eqn:eqnnoname1} is an \emph{Anyonic Symmetry} (AS) of the fully
 braided fusion category obtained by deconfining the $h/2e$ flux vortices as well. This
 means that the AS $\m\rightarrow\m^p$ preserves the
 braiding of the qps (in both the original theory and the new
 theory with $h/2e$ deconfined).\footnote{Actually the
constraint $\delta_{\mathcal{E}}=2^2\delta_{\bf m}$ (mod $2\pi$) does not
necessarily pin down $\delta_{\bf m}=\frac{\pi}{8n+4}$. For instance, it could
also be $\delta_{\bf m}=\frac{(4n+3)\pi}{8n+4}$. However, this ambiguity does
not matter in the end because we can always reshuffle our ${\bf
m}$ so that the new ${\bf m}$ has $\delta_{\bf m}=\frac{\pi}{8n+4}$. To see
this, suppose there are two integers $p_1$ and $p_2$ so that $p^2_i\equiv1$ mod
$16n+8$ and $\delta_{\mathcal{E}}=2^2\delta_{p_i}$ These two integers are
related by $p_1\equiv p_2(p_1p_2)$ and $p_2\equiv p_1(p_1p_2)$, and the
multiplication by $p_1p_2$ is always an AS.} We will discuss AS in greater detail in
 Sec. \ref{sec:AS}.

%
%
%
Let us mention one more important aside.   The Gauss-Milgram
 formula~\cite{FrohlichGabbiani90,Kitaev06} relates the chiral central charge
 $c_-$ -- a quantity that dictates the thermal Hall conductance -- to the
 statistics of the qps. Before adding the superconductor each Laughlin state
 has $c_{-}=1.$ This same relationship is also satisfied in the theory with the
 augmented TO if we include the full set of $8n+4$ qps: \begin{align} e^{i\pi
	 c_-/4}=\sum_{\bf{a}}\theta_{\bf{a}}=\sum_{j=0}^{8n+3}e^{i\frac{\pi}{4}\frac{j^2}{2n+1}}=e^{i\pi/4}
 \end{align}  Thus, with gauged $\mathbb{Z}_2$ fermion parity symmetry, the SC-FCI has an extended {\em bosonic} TO which can be described by a
 $U(1)$-Chern-Simons theory at level $4n+2$ with $K_{\mbox{sc}}=8n+4 $. We
 also note that this theory ensures that the electron $\psi$ (qp vector 4n+2)
always braids with a phase of $-1$ around $\m$ (qp. vector 1), and the central charge of the edge theory is unmodified. 
 
Now let us consider some explicit solutions to Eq. \eqref{eqn:eqnnoname1}.
Remarkably, Eq. \eqref{eqn:eqnnoname1} is always satisfied irrespective
of $n$ for the two choices $\m^{a}\rightarrow\m^{-a}\equiv\m^{8n+4-a}$ or $\m^{a}\rightarrow \m^{a}\times\psi^{a}=\m^{a+4n+2}$. 
Hence, these are quite general anyonic symmetries, and physically they correspond to  {\em charge conjugation} 
and the aforementioned {\em fermion parity flip} symmetry respectively. i) Charge
conjugation $\m^{a}\rightarrow\m^{-a}$ is well-known and transmutes qps into
quasiholes and vice-versa. ii) On the other hand, the fermion parity flip symmetry
 is the main subject of
this paper, and corresponds to an even (odd) number of fermions $\psi=\m^{4n+2}$ being pumped into a vortex with even (odd) vorticity. 
For example, the action of the symmetry changes the local fermion parity
of a single vortex $\m,$ and since  $\psi^2=1$, the same is true for any $\m^{2p+1}$ for integer $p.$ Even-vorticity objects retain their fermion parity during such a process.
Also, since adding a fermion twice does nothing (modulo a Cooper pair) this symmetry generates a $\mathbb{Z}_2$ group.
 iii) A third AS  corresponds to a composition of these
 symmetries and is also always present in fermionic Laughlin states and corresponds to $\m^{a}\rightarrow
 \left(\m^{4n+2+a}\right)^{-1}=\m^{-(4n+2+a)}\equiv \m^{4n+2-a}$.

In particular, for the Laughlin $\nu=1/3$ state with $n=1$ the theory has $12$ qps, all of which are generated by $\m.$ These anyonic symmetries translate to
i) $\m\rightarrow \m^{11}$ (charge conjugation), ii) $\m\rightarrow \m^7$(Fermion
parity flip), and iii) $\m\rightarrow \m^5$ (composition of conjugation and parity
flip) respectively. Since  all qps are generated by $\m$ for the Laughlin series, one can determine the action of the AS on the rest of the qps from the relations above. We caution that for a general Laughlin state with arbitrary $n$ there
can be more anyonic symmetries which obey equation \eqref{eqn:eqnnoname1}
than these three mentioned above. These additional AS are not generic and must be determined on a case by case basis.

\subsection{Description of defects in Laughlin states} 
\begin{figure}[htbp] \begin{center}
\includegraphics[width=0.4\textwidth]{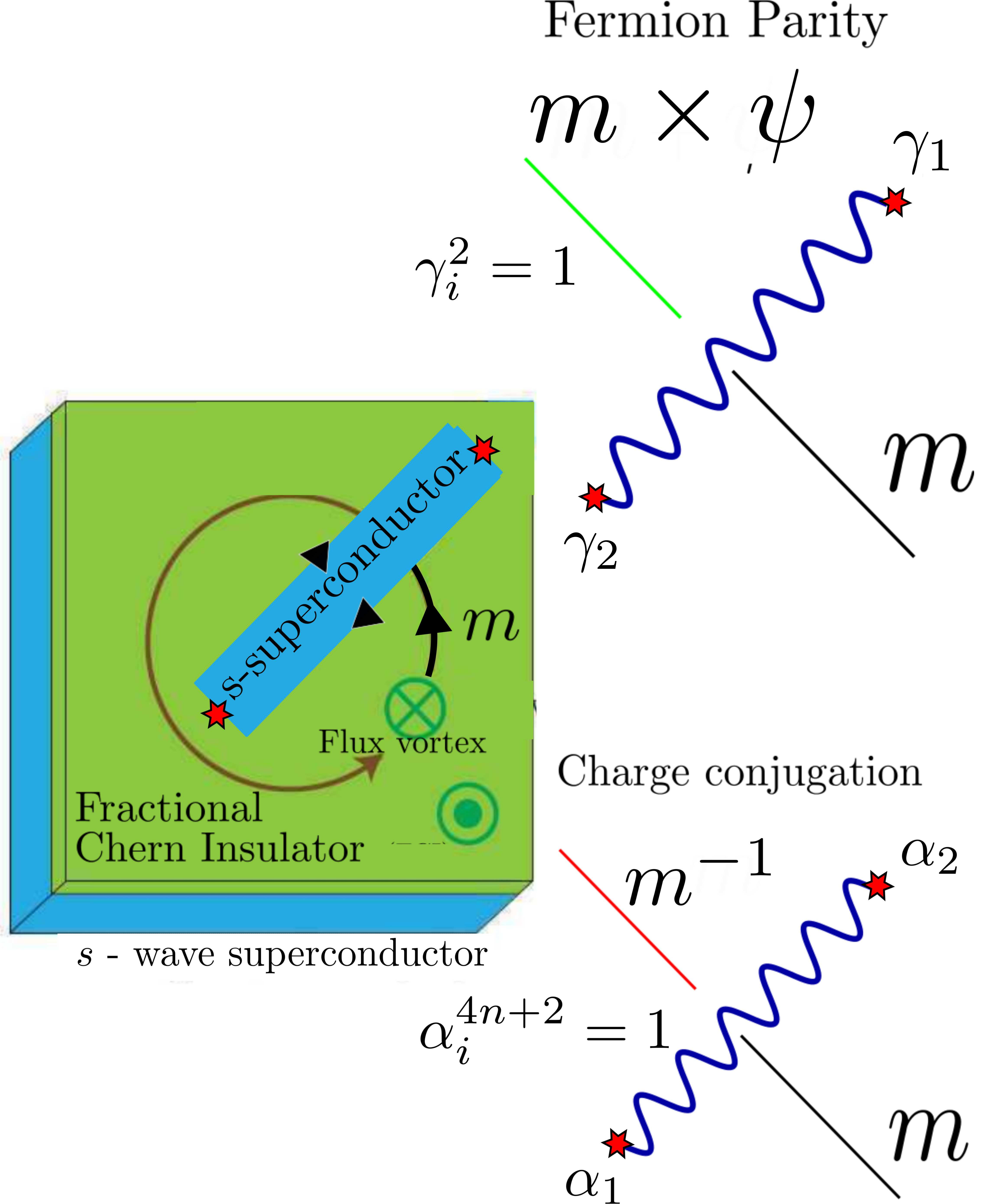} 
\end{center}
\caption{(color online) Layered heterostructure of a Laughlin state and an s-wave 
superconductor. Blue superconducting substrate with green FCI layer. Counterpropagating edges on the substrate
are shown by the black arrows. A flux vortex $\m$ orbiting around the twist defect (red stars)
gets mutated by the anyon permutation symmetry of the system. The two AS considered are shown schematically
with the superconducting trench behaving like a branch cut. 
MBS $\gamma_1$ and $\gamma_2$ at the
edges of the trench which realizes fermion parity flip. When, charge
conjugation symmetry is realized the corresponding bound states $\alpha_{1/2}$ are
parafermionic.} \label{fig:SCTOlayer} 
\end{figure}

In this subsection we will give a heuristic description of the twist defects corresponding to the AS operations mentioned above. For a more technical discussion see Sec. \ref{sec:twistdefect}. To develop intuition about the non-Abelian nature of twist defects in our context, consider
the geometry in Figure \ref{fig:SCTOlayer}. A (blue) s-wave superconducting
substrate lies
beneath the (green) FCI layer. Additionally, there is a trench hollowed out in the FCI layer with 
counter-propagating edge modes as shown by the grey arrows. \footnote{We consider the FCI here for convenience so that we do not have to worry about spatial-dependent g-factors.} One can include backscattering terms
between the edges which correspond to tunneling terms across the trench. The net effect of these terms is to gap 
out the edge modes,  however, these tunneling terms can be such that they permute the bulk anyon qps by the action of an AS
as they tunnel across the trench.\cite{BarkeshliJianQi,khan2014,Kitaev06,Bombin,Levin13,Bombin,YouWen,TeoRoyXiao13long}. We will demonstrate this explicitly in Section \ref{sec:AS}.
The two AS mentioned in the previous subsection, charge conjugation
and fermion parity flip, are shown in the diagram \ref{fig:SCTOlayer}. The trench acts  
like a branch
cut  and the ends play the role of twist defect cores (shown by the red stars).

We can argue for the quantum dimension of the NAZM on a twist defect core as follows. The charge $Q$ of the superconducting trench is defined modulo $2e$. However, $Q$ can change during a
 qp tunneling process. First,  let us suppose the edge couplings are such that the trench enacts the charge conjugation operation as qps tunnel across. A single
 particle $\m$ tuneling across the trench changes $Q$ by $2q_{\m}=e\nu=e/(2n+1)$ (recall that $\m$ is the qp with the smallest charge in the system $q_{\m}=e\nu/2=\frac{e}{4n+2}$). Thus, it is possible for $Q$ to change in
 increments of $e/(2n+1)$ mod $2e$. This leads to a ground state degeneracy of $4n+2$ labelled by the different values of $Q$,
 $0,\frac{e}{2n+1},\frac{2e}{2n+1},\cdots,2e.$. The increase in ground state degeneracy can be attributed to the presence of the two NAZM at the ends of the trench which are generalized  
 $\mathbb{Z}_{4n+2}$ MBS/parafermion
 twist defects with quantum dimension $\sqrt{4n+2}$. 
 
 On the other hand, we can let the edge couplings enact the fermion parity flip symmetry operation as qps tunnel across. Now, however the charge of
 the trench can only change in increments of $e$. Thus, there are just $2$ ground states of the system which are labelled by $Q=0,e$.
 This implies the existence to two NAZMs at the ends of the trench that are simply MBS with quantum dimension 
 $\sqrt{2}$.  To see this another way, let us consider
 dragging $\m$ around one end of the trench. 
As $\m$ morphs into $\m\times\psi$ the ground state of the trench switches between the two degenerate sectors
and indicates a change in its local fermion parity. This transition on the trench is naturally accompanied by a level crossing among the Caroli-de Gennes-Matricon bound states in the vortex core. Indeed, this was verified numerically for a Chern insulator with unit Chern number placed in contact 
with an s-wave superconductor in Ref. \onlinecite{khan2014}. 

From a simple analysis we can verify that the total fermion parity of the system is conserved.
To see this let us consider the two MBS $\gamma_1$ and $\gamma_2$ located at the two ends of the trench.
Together, they form a complex fermion $c=\gamma_1+i\gamma_2$.  As the $h/2e$ flux ($\m$ qp) is dragged around $\gamma_2$,
the phase of the superconducting order parameter (bilinear in fermion
operators) winds by $2\pi$, and hence the enclosed Majorana fermion picks up a
phase of $\pi$ and changes sign:
$\gamma_2\rightarrow -\gamma_2$.
The fermion $c$ now gets conjugated 
$c=\gamma_1+i\gamma_2\rightarrow\gamma_1-i\gamma_2=c^{\dagger}$. Thus, the change in fermion parity 
in the vortex is compensated by the change in fermion parity encoded in the two
MBS twist defects at the ends of the trench. In fact, $c\rightarrow c^{\dagger}$ encodes the change in total fermion
parity of the superconducting trench and just reflects the addition/removal of
the electron from it, i.e., the switch in the value of $Q$ for the trench ground state. 

We have now completed a thorough discussion of the Laughlin states. For more examples of Abelian, fermionic FQH states with gauged fermion parity we point the reader to Appendix \ref{sec:examples}. In the following sections we present the general form of the theory beginning with gauging fermion parity and then moving on to generic AS and the associated twist defects. 

\section{Gauging the fermion parity of an Abelian TO
state}\label{sec:gaugingfermionparity}
In the previous section we have seen an important pattern:
\begin{enumerate}[i)]
 \item A Laughlin state in proximity to an s-wave superconductor has the same
	 \emph{fusion} structure as the more exotic system with
	 extended TO where the
 fermion parity is gauged and the $h/2e$ flux vortices are deconfined.
 \item The twist defects associated to an anyonic symmetry of the fermion parity gauged theory can also exist in the original
proximity-coupled Laughlin state system.
\end{enumerate}
Interestingly,  this pattern applies for general Abelian states built from local electrons.
 To show this, we will first develop the effective topological field theory for
a fermion parity gauged Abelian TO state. Similar analysis has been done to
couple a fermionic state to a $\mathbb{Z}_2$ gauge field to obtain a bosonic 
theory\cite{YouBIalexchengxu}, and in classifying symmetry enriched phases\cite{lu2013classification}. After we have the effective theory for the gauged TO system, we will analyze it to determine the AS and corresponding twist defects that can appear in the ungaged theory, and hence are relevant for experimental realizations of superconductor-TO heterostructures. 

A general Abelian TO state is described by the bulk Chern Simons action
\begin{align}
\mathcal{L}_{\text{bulk}}&=\frac{K_{IJ}}{4\pi}\epsilon^{\mu\nu\lambda}\alpha_{I\mu}\partial_{\nu}\alpha
_{J\lambda}-\frac{1}{2\pi}t_{I}\epsilon^{\lambda\mu\nu}A_{\lambda}\partial_{\mu}\alpha_{I\nu},\nonumber\\
&\quad I,J=1,2,\cdots,N. 
\label{eq:bulkCS}
\end{align}
Here, $\bf{t}$ is the charge vector which determines how the external electromagnetic gauge field $A$
couples to the gauge fields $\alpha_I$.


If we treat the s-SC as being TO\cite{HanssonOganesyanSondhi04,BondersonNayak13}, then the $h/2e$ vortex $\m$ is deconfined.
This represents the deconfined phase of a $\mathbb{Z}_2$ gauge theory, and there are four distinct qp sectors: $1$ (the superconducting vacuum condensate), 
${\bf m}$ (the bosonic $h/2e$ vortex), 
$\bb{\psi}$ (the fermionic Bogoliubov-de Gennes quasiparticle),
and ${\bf e}$ an excited Caroli-de Gennes-Matricon vortex state, ${\bf e}=\m\times\psi$.
The TO is identical to the toric code model\cite{Kitaev03}, and is
captured by a  two-component Chern-Simons theory with
$K$ matrix $K_{\text{s-sc}}=2\sigma_x$ in the basis $(a,b)^T$ of $U(1)$ gauge fields.
The corresponding topological part of the action takes the form
\begin{align}
\mathcal{L}_{\text{s-sc}}&=\frac{1}{\pi}\epsilon_{\mu\nu\lambda}a^{\mu}\partial^{\nu}b^{\lambda}
\label{eqn:TopSsc}
\end{align}
and $\m$  and $\bb{e}$ carry unit charge under $b$ and $a$ respectively. In this basis of gauge fields, the charges carried by ${\bf e},{\bf
m},\psi$ are  $(1,0),(0,1),$ and $(1,1)$ respectively.  By itself, this TO
carries an electric-magnetic anyonic symmetry, ${\bf e}\leftrightarrow{\bf m}$.
It effectively interchanges fluxes with \emph{opposite fermion parities}, while keeping
the BdG fermion $\psi$ invariant. This permutation of the qp sectors leaves the topological
information--the spin and braiding statistics--unchanged and, hence is  an
AS of the theory.

Now we can put the two pieces together to find:
\begin{align}
\mathcal{L_{\text{sc}}}=\frac{K_{IJ}}{4\pi}\alpha_I \wedge d\alpha_J
-\frac{t_I}{2\pi}a\wedge d\alpha_I+\frac{1}{\pi}a\wedge db
\label{eqn:eqn1} 
\end{align}
where $A$ has now been replaced by the dynamical $U(1)$ gauge field
 $a,$ since we are treating the s-SC as having dynamical gauge fluctuations.  Integrating out $a$ leads to the
constraint (neglecting large gauge transformations) 
\begin{align}
2b=t_I\alpha_I. \label{eqn:gaugeconstraint} 
\end{align}
This constraint encodes the interplay between the superconducting flux and the quantum Hall effect; in particular, it is just an alternative statement of $\mathcal{E}=\m^2.$
To see this we note that $\m^2$ carries a charge of $2$ under $b,$ while the $h/e$ flux $\mathcal{E}$ is described by the charge vector
$\bb{t}$ in the basis of gauge fields $\alpha$. Hence,  we see that the
effective Lagrangian of the fermion parity gauged theory has precisely  captured
the fusion rule $\mathcal{E}=\m^2$ that we expect to hold true for any Abelian
TO system. Fusion for the fermion parity gauged theory is hence just the original fusion
theory of Eq. \ref{eq:bulkCS} augmented by the fusion rule
$\mathcal{E}=\m^2$. Thus, the fusion structures for both the fermion parity
gauged system (which is fully braided) and the original system plus the inclusion of the semiclassical $h/2e$ fluxes
(partially braided), are identical. 

Thus, the main result of this Section is that the effective theory is obtained via Eq. \ref{eqn:eqn1} with the constraint Eq. 
\ref{eqn:gaugeconstraint}. Alternatively, the emergence of this new theory can be
understood using ideas of stable equivalence as we show in Appendix
\ref{sec:Laughlinstabequiv}. Let us now comment on a number of properties
that our fermion parity gauged theory obeys in the following subsection. 
\subsection{General Properties of the Gauged Theory}\label{sec:physicalconstraints}
Let us consider a state at filling $\nu$. By Laughlin's argument, the $h/e$ flux $\mathcal{E}$
has charge $q_{\mathcal{E}}=\nu$ and statistical angle $\delta_{\mathcal{E}}=\pi\nu$.
As our contraint implies,  $\m^2=\mathcal{E}$ in general, and we have $q_{\m}=\nu/2$ and $\delta_{\m}=\pi\nu/4$.
Also, the electron $\psi$ should naturally obtain a -1 statistical phase when it braids around an $\m$.

Now let us consider the particle content of the gauged theory. All qps in the original Abelian theory have quantum dimension $d=1$ by definition.
Since the new qps in the fermion parity gauged theory are formed by fusion of $\m$ with the original excitations, 
then all qps in the new theory also have $d=1$, and hence the TO will be Abelian.
Using general results from Refs. \onlinecite{barkeshli2014,Teotwistliquids} which relate the
total quantum dimension of a system before and after a discrete symmetry is gauged we should have that
\begin{align} \mathcal{D}_{\text{gauged}}&=\mathcal{D}_0|G|
\label{eqn:Qdimensiongauged} \end{align} where $\mathcal{D}$ denotes
the total quantum dimension of the topological phase  determined as
$\mathcal{D}=\sqrt{\sum_{i}d_{i}^2}$, $\mathcal{D}_0$ denotes the quantum dimension of the original theory, and
$|G|$ is the order of the discrete symmetry group being gauged. For our case
$|\mathbb{Z}_2|=2$ and $d=1$ for
all qps in the original as well as the gauged theory. Hence, using
Eq. \eqref{eqn:Qdimensiongauged} \begin{align}
\mathcal{D}_{\text{SC}}&=2\mathcal{D}_0\nonumber\\ d_{i}=1
\forall i &\implies |\mathcal{A}_{\text{SC}}|=4|\mathcal{A}_0|
\label{eqn:degengauged} \end{align}
where $\mathcal{A}$ is the set of topologically distinct qps in the theory.
Thus, the total number of quasiparticles always increases 4 fold.
In fact, we already saw this for the
Laughlin state in the previous section where the number of qps  increased from $2n+1$
to $8n+4$.

Also, we should find that during the gauging process the chiral central charge $c_{-}$ which determines the number of chiral
 edge modes in the system should not change.

 In summary,
\begin{align}
e^{i\theta_{\bb{\psi,m}}}=-1;\quad & q_{\bb{m}}=\nu /2\nonumber\\
\delta_{\bb{m}}=\pi\nu/4;\quad &|\mathcal{A}_{\text{SC}}|=4|\mathcal{A}_0| \nonumber\\
c_- =c_{-,\text{SC}}&
\label{eqn:conditionhalfflux} 
\end{align}
Let us see how these considerations hold for the Laughlin states. 

\subsection{Laughlin States} 
\label{sec:Laughlinsection} 
Laughlin states at filling $\nu=\frac{1}{2n+1}$ have the action \begin{align*}
\mathcal{L}&=\frac{2n+1}{4\pi}\epsilon^{\mu\nu\lambda}\alpha_{\mu}\partial_{\nu}\alpha_{\lambda}-
\frac{1}{2\pi}
\epsilon^{\mu\nu\lambda}A_{\mu}\partial_{\nu}\alpha_{\lambda}
\end{align*} before coupling to the s-SC. Here $K=2n+1$ and the charge vector
${\bf{t}}=1$. When gauging the fermion parity symmetry we need to find the solution to Eq. \eqref{eqn:gaugeconstraint}. In this
$1$ component case this  is simply $2b=\alpha$.   Re-expressing the theory in terms of $b$ yields \begin{align*}
\mathcal{L}&=
\frac{8n+4}{4\pi}\epsilon^{\mu\nu\lambda}b_{\mu}\partial_{\nu}b_{\lambda}.
\end{align*} Thus the new state is characterized by
$K_{\mbox{sc}}=8n+4$.
Using $2b=\alpha$, we effectively have $\bb{t_{\mbox{sc}}}=2$.
The presence of the superconductor implies that charges are
conserved modulo 2. Hence, the charge vector $\bb{t_{\mbox{sc}}}$ is also defined
modulo 2 and t the topological properties of the qps are
invariant under the addition of a charge $2e$ Cooper pair.

This theory satisfies all the constraints in
\ref{sec:physicalconstraints}:
\begin{align*}
\delta_{\m}&=\pi/(8n+4)=\pi\nu/4\nonumber\\
c_-&=1\quad \mbox{(chiral central charge is preserved).} \end{align*}

\section{Fermion parity flip anyonic symmetry in a general abelian
state}\label{sec:fermionparitygeneral}
Previously we have seen that the topologically ordered phase of the
s-SC when fermion parity is gauged  has an electromagnetic AS
$\bb{m}\leftrightarrow \m\times\psi=\bb{e}$. 
A more general version  of this symmetry, which connects qps having different fermion parity,
exists in a general fermionic TO state in contact with an s-SC. We will now demonstrate this.

For convenience let us define the {\em vorticity} of
a qp $\bb{x}$ as the charge of $\bb{x}$ under the gauge field $b$. Thus,
$\m$ has a charge of $1$, the $h/e$ flux has a charge of $2,$ and so on. 
We will denote a qp $\bb{x}$ with vorticity $p$  as $\bb{x}_p$. Physically, the
vorticity just counts the number of $\m$ excitations in the qp.
The general definition of the fermion parity flip anyonic symmetry is
\begin{align}
\bb{x}_{p}&\rightarrow \bb{x}_{p}\times\psi^p. \label{eqn:fermionparityflipas}
\end{align} In particular, $\m\rightarrow \m\times\psi,$ and since 
$\psi^2\equiv 1$, i.e., the vacuum then 
\begin{align}
\bb{x}_{p}\rightarrow &\bb{x}_{p} \mbox{ for even $p$}\nonumber\\
\bb{x}_{p}\rightarrow &\bb{x}_{p}\times \psi \mbox{ for odd $p$.}
\label{eqn:fermionparityflipas1}
\end{align}

To show this is an AS we need to analyze the braiding phases, which we will do using the ribbon formula \cite{Kitaev06}.
\begin{align}e^{i\delta_{\bf
	z}}=\vcenter{\hbox{\includegraphics[width=0.5in]{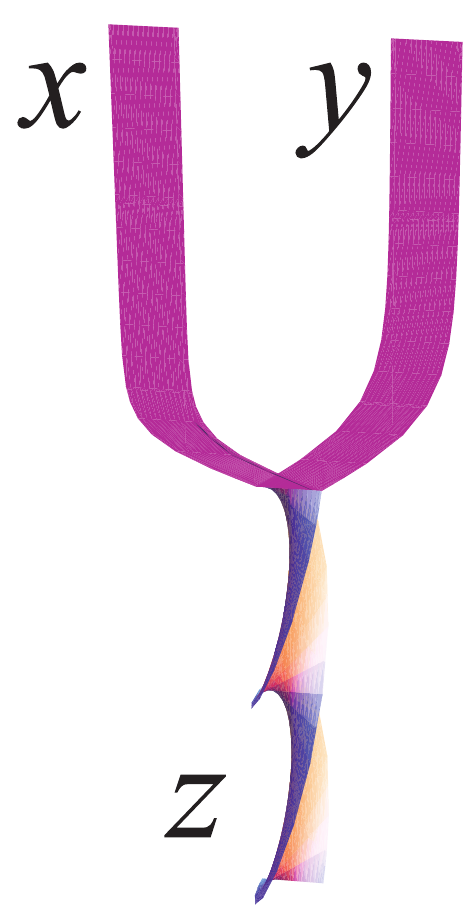}}}=\vcenter{\hbox{\includegraphics[width=0.5in]{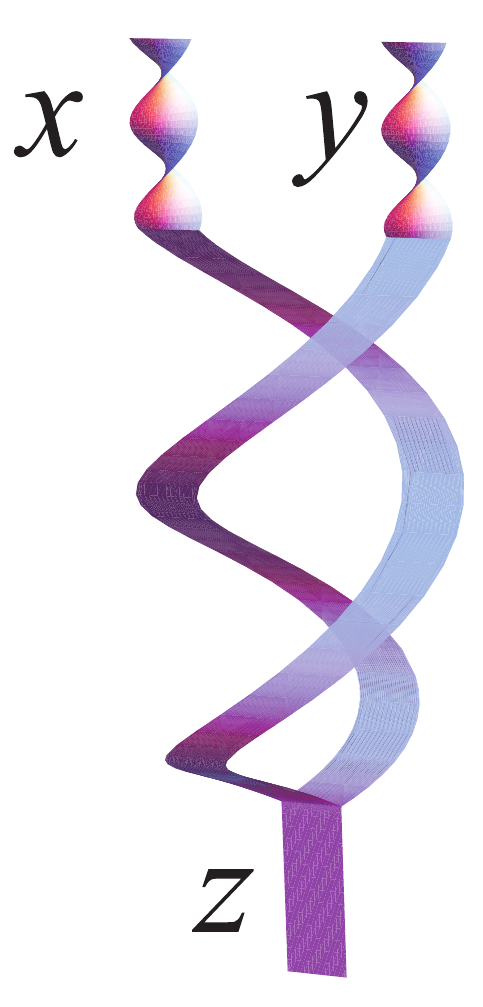}}}=e^{i\theta_{\bf
	xy}^{\bf z}}e^{i\delta_{\bf x}}e^{i\delta_{\bf
	y}}\label{ribbonapp}\end{align} where $e^{i\theta_{\bf xy}^{\bf z}}$ is
	the gauge-independent ($2\pi$) monodromy phase between ${\bf x}$ and
	${\bf y}$ with a fixed overall fusion channel ${\bf z}$ and
	$\delta_{\bf{z}}$ is the exchange phase of qp $\mathbf{z}$.
Indeed, we will show that the modular $S$ and $T$ matrices are left invariant 
under fermion parity flip symmetry in Eq. \ref{eqn:fermionparityflipas1}. The $S$ matrix can be written as 
\begin{align}
	S_{\mathbf{x,y}}&=
	\frac{1}{\mathcal{D}}\sum_{\mathbf{z}}N^{\mathbf{z}}_{\mathbf{xy}}\frac{e^{i\delta_{\mathbf{z}}}}
	{e^{i\delta_{\mathbf{x}}}e^{i\delta_{\mathbf{y}}}}d_\mathbf{z}
\label{eqn:kitaevribbon}
\end{align}
where $N^{\mathbf{z}}_{\mathbf{xy}}$ is defined by
the fusion rule $\bb{x\times y}=N^{\mathbf{z}}_{\mathbf{xy}}\bb{z}$.
For Abelian phases $d_\mathbf{z}=1$ and $N^{\mathbf{z}}_{\mathbf{x{y}}}=1$ only when $\mathbf{z=x\times{y}}$, otherwise it is
$0$. Thus for an Abelian system this simplifies to
\begin{align}
	S_{\mathbf{x,y}}&=\frac{1}{\mathcal{D}}\frac{e^{i\delta_{\mathbf{x{y}}}}}{e^{i\delta_{\mathbf{x}}}e^{i\delta_{\mathbf{y}}}}.
\end{align}

To reduce clutter we will assume $\bb{x}$ has vorticity $p$ and $\bb{y}$ has vorticity $q$
without explicitly indicating them
via subscripts. We note that vorticity is additive under fusion and in the expressions below,
$\theta_{\mathbf{a,b}}$ is the braiding phase of qp $\mathbf{a}$ around
$\bb{b}$.
Under the fermion parity flip
symmetry
\begin{align}
	e^{i\delta_{\mathcal{P}\bb{x}}}=e^{i\delta_{\bb{x}\psi^p}}&=e^{i\delta_{\bb{x}}}{(-1)}^pe^{i\theta_{\bb{x},\psi^p}}\nonumber\\
	(\mbox{As $\bb{x}$ has vorticity p)\quad}
	e^{i\theta_{\bb{x},\psi^p}}&=e^{i\theta_{\bb{m}^p,\psi^p}}={(-1)}^{p^2}=(-1)^p\nonumber\\
	\implies e^{i\delta_{\mathcal{P}\bb{x}}}&=e^{i\delta_{\bb{x}}}{(-1)}^p{(-1)}^p=e^{i\delta_{\bb{x}}}.
\label{eqn:exchangestatASgeneral}
\end{align}
Thus the exchange statistics and hence the $T$ matrix is preserved.
Proving invariance of braiding is now straightforward:
\begin{align}
S_{\mathbf{\mathcal{P} x,\mathcal{P}
y}}&=\frac{1}{\mathcal{D}}\frac{e^{i\delta_{\mathbf{\mathcal{P} x\mathcal{P}
y}}}}{e^{i\delta_{\mathbf{\mathcal{P} x}}}e^{i\delta_{\mathbf{\mathcal{P}
y}}}}= \frac{1}{\mathcal{D}}\frac{e^{i\delta_{\mathbf{\mathcal{P} (x
y)}}}}{e^{i\delta_{\mathbf{\mathcal{P} x}}}e^{i\delta_{\mathbf{\mathcal{P}
y}}}}\nonumber\\
&=\frac{1}{\mathcal{D}}\frac{e^{i\delta_{\mathbf{x{y}}}}}{e^{i\delta_{\mathbf{x}}}e^{i\delta_{\mathbf{y}}}}.
\label{eqn:braidinginvariant}
\end{align}
Thus, under very general conditions we have shown that our fermion parity flip symmetry is an AS. This is one of the primary results of this article. Now that we have such a symmetry we will try to use it to generate possible twist defects. 

\section{Realizing anyon permutation at a gapped interface}\label{sec:AS}
In this section we briefly review \cite{khan2014,Haldane95,Levin13,BarkeshliJianQi13long} for completeness how an anyon permutation symmetry can be realized at the 
gapped  interface between two edges of a TO medium. This interface region is comprised of counter-propagating edge states on both sides of the interface, and tunneling processes across the interface. Since there is a natural connection between the K-matrix bulk theory and an edge theory, let us first present how an AS in Abelian TO states can be understood within
the $K$-matrix formalism, and then we will connect this back to the physics at a quasi-1D interface. At that point we will give an example that shows how the various different AS, including the fermion parity flip, can be realized in a $\nu=1/3$ state. 

It is well known that the choice of a $K$-matrix that describes an Abelian TO is not unique, but
is only determined up to $GL(N,\mathbb Z)$ basis transformations. Thus, $K\rightarrow WKW^T$ $W\in
GL(N,\mathbb Z) (\vert\det W\vert =1)$ where the quasiparticles $[\mathbf{l}]$  $\in \mathbb{Z}^N,$
are transformed to $[W\mathbf l]$, leaves the physical content of the theory invariant. Braiding and exchange are determined
by the $K$-matrix alone, and these transformations must leave the topological properties of the
quasiparticles unchanged. Thus,
$S_{\mathbf a,\mathbf b}=S_{\mathbf Wa,W\mathbf b}$, $\delta_{\mathbf{a}}=\delta_{W\mathbf{a}}$.

The subset of all transformations $W$ which leave the K-matrix identically unchanged form the group of automorphisms of the $K$
matrix
\begin{align}\mbox{Aut}(K)=\left\{W\in	GL(N;\mathbb{Z}):WKW^T=K\right\}\label{eqn:Aut(K)}.\end{align}
Some $W$ act trivially on the anyon labels $\mathbf a$, i.e.,  they preserve
the anyon vector up to a local particle addition of the form $K\mathbb{Z}^N$.
This set forms a normal subgroup $\mbox{Inner}(K)$ of inner
automorphisms
\begin{align}\mbox{Inner}(K)=\left\{W_0\in\mbox{Aut(K)}:[W_0{\bf a}]=[{\bf
a}]=\mathbf a+ K\mathbb Z^N\right\}.
\end{align}
Instead, we are interested in the {\em{non trivial}} anyon relabelings that represent anyonic symmetries. These
form the set of outer automorphisms
\begin{align}\mbox{Outer}(K)=\frac{\mbox{Aut}(K)}{\mbox{Inner}(K)}.\label{outerdef}\end{align}
Thus, $\mbox{Outer}(K)$ is the AS group of the Abelian
topological phase characterized by $K$.

Unfortunately, this classification misses some possible AS operations because we have ignored another equivalence, i.e., stable equivalence\cite{cano2013bulk,Levin13,BarkeshliJianQi13long,lu13}. Stable equivalence is the statement that a K-matrix is equivalent to another K-matrix if they differ only by the addition of a trivial, decoupled sector (one might argue about the definition of ``trivial", but we will not worry about that for now). Thus, while each element of  $\text{Outer}(K)$ represents a possible AS, it is not always sufficient to capture all of the AS of a given set of anyons. 
This insufficiency is essentially true because the braiding and exchange statistics $\theta$ and $\delta$ that must be preserved by an AS are complex phases and
are only defined modulo $2\pi$. Thus, an AS 
only needs to preserve 
the scaling dimension modulo integers.

Interestingly, it appears that given a particular AS, one can find a representative element for each AS as an outer automorphism, but it sometimes becomes
necessary to consider an enlarged $K$-matrix which is stably
equivalent to the
original one. In fact, one could argue from the results of Ref. \onlinecite{cano2013bulk}, although it has not been precisely proven, that we need only add an
additional $2\times 2$ matrix and consider $K\oplus\sigma_z$ or
$K\oplus\sigma_x$ (corresponding to adding topologically trivial fermionic or
bosonic modes to the system) to realize all anyonic symmetries $\mathcal{P}$ using $W$
matrices as outlined in equations \eqref{eqn:Aut(K)},\eqref{outerdef}. 
With this in mind, let us consider the $\nu=1/3$ Laughlin state  represented by $K$=3. We have
already seen that when placed in proximity to an s-SC and deconfining the $\m$ qp, the emergent TO phase is the bosonic state $K=12$. We have also discussed that this theory has 3
AS $\m\rightarrow \m^5,\m^7,$ or $\m^{11}$ respectively. In the $1$ component picture, the only
AS realizable is $\m\rightarrow m^{-1}\equiv \m^{11}$, corresponding to $W=-1$. Thus, to realize the other two AS,
we must enlarge the $K$ matrix to, e.g., $12\oplus\sigma_x$ as is appropriate for bosonic
states. Now we see that all the AS can be realized, in particular
\begin{align}
	W_7=\left(
\begin{array}{ccc}
 7 & 12 & -24 \\
 1 & 2 & -3 \\
 -2 & -3 & 8 \\
\end{array}
\right);W_5=\left(
\begin{array}{ccc}
 5 & -12 & 12 \\
 1 & -2 & 3 \\
 -1 & 3 & -2 \\
\end{array}
\right)	
\label{eqn:1/3rdstablequiv}
\end{align}
realize $\m^7$ and $\m^5$ respectively.
These non trivial outer automorphisms are realized
up to inner automorphisms,  thus,
two possible realizations for $\m\rightarrow\m^{11}$ are
\begin{align}
	-\mathbbm{1}_3 \quad \mbox{and} \left(
\begin{array}{ccc}
 -1 & 0 & 12 \\
 -1 & -1 & 6 \\
 0 & 0 & -1 \\
\end{array}
\right) 
\label{eqn:m11twomatrices}
\end{align}
\begin{figure}[htbp] \begin{center}
\includegraphics[width=0.4\textwidth]{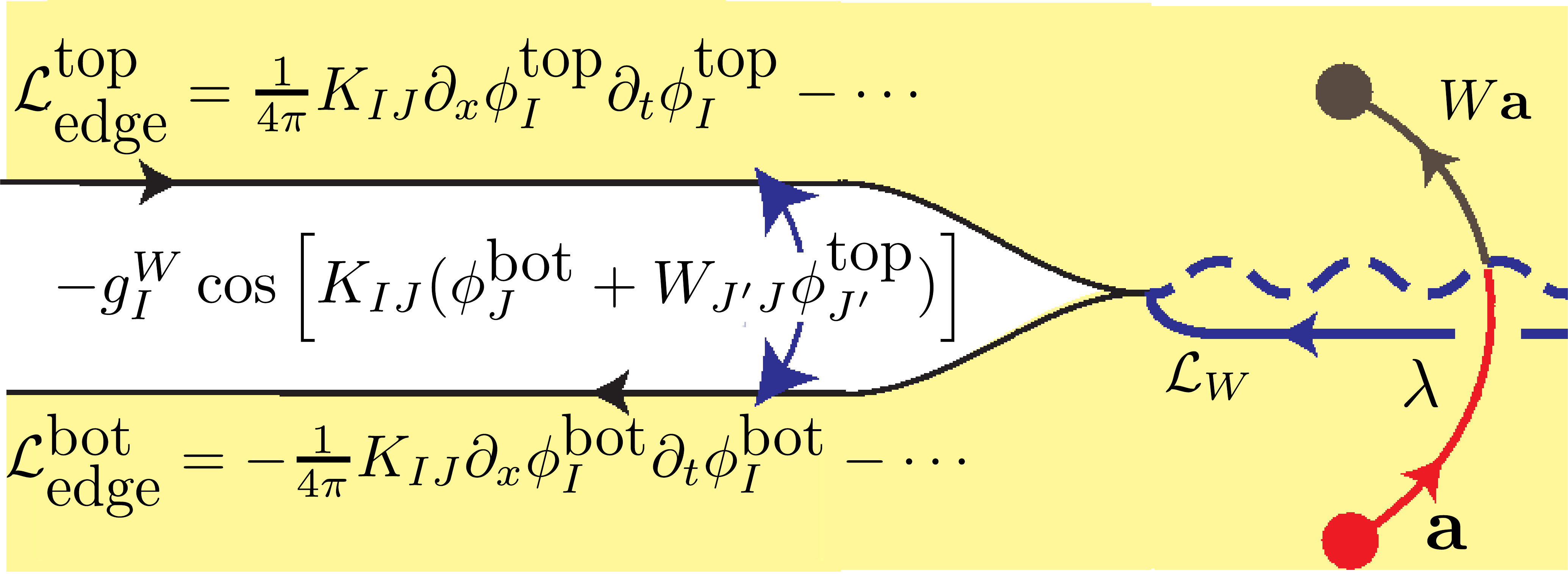} \end{center}
\caption{(color online) Reproduced from Ref. \onlinecite{khan2014}. The term 
$\mathcal{L}_W$ gaps out counterpropagating edge modes and mutates qp $\mathbf{a}$ to
$W\mathbf{a}$.
} \label{fig:edgecouplings}
\end{figure}

Now let us consider a one-dimensional interface between two identical
TO states as in Figs. \ref{fig:SCTOlayer} and \ref{fig:edgecouplings}. The realizations for the AS above are important because they will enter our construction. 
There are counter propagating edge
modes are represented via
\begin{align}\mathcal{L}^{top}_{edge}+\mathcal{L}^{bottom}_{edge}=\frac{1}{4\pi}
K^{\sigma\sigma'}_{IJ}\partial_x\phi_I^\sigma\partial_t\phi_J^{\sigma'}{
+\frac{1}{2\pi}t^{\sigma}_{I}\epsilon^{\mu\nu}\partial_{\mu}\phi^{\
\sigma}_IA_{\nu}}  \label{interfaceK}\end{align} 
where $\sigma=0,1=R,L$
labels right and left moving modes, $\phi^R_I$ ($\phi^L_I$) are the boson fields living along
the top (bottom) edge, and
$K^{\sigma\sigma'}_{IJ}=(-1)^\sigma\delta^{\sigma\sigma'}K_{IJ}$.

Corresponding to every element $W$ of the AS group we can write down a gapping term for the
edge
\begin{align}\delta\mathcal{L}_W=- g^{W}_I\cos\left[K_{IJ}
\left(\phi^L_J+W_ {
J'J }\phi^R_{J'}\right)\right].\label{edgecoupling}
\end{align} This term represents local boson tunneling between $e^{-iK_{IJ}\phi^L_J}$ and
$e^{iK_{IJ}W_{J'J}\phi^R_{J'}}$ during which an anyon $\mathbf a$ transforms to $W\mathbf
a$ as it crosses the interface (for a bulk qp $\mathbf a$ the vertex
operators on the left and right edges are $\psi_{\mathbf{a}}^L=e^{-i\mathbf{a}.\mathbf\phi^L}$ 
and  $\psi_{\mathbf{a}}^R=e^{i\mathbf{a}.\mathbf\phi^R}$ respectively in our convention).
The gapped interface is characterized by the vacuum expectation value 
\begin{align}\langle\phi^L_I+W_{JI}\phi^R_J\rangle=2\pi(K^{-1})_{
IJ } \lambda_J , \quad\mbox{for ${\bf
\lambda}\in\mathbb{Z}^N$}\label{eq:bosonpinning}\end{align}
or alternatively,
\begin{align}
\left\langle\left(\psi^{W\mathbf{a}}_R\right)^\dagger\psi^{\bf a}_L\right\rangle=\langle e^{-i({\bf
a}\cdot\boldsymbol{\phi}^L+(W{\bf a})\cdot\boldsymbol{\phi}^R)}\rangle=e^{-2\pi i{\bf
a}^TK^{-1}{\bf \lambda}}.\label{condensate}
\end{align}\noindent
This represents $\mathbf a$ transforming into $W\mathbf a$ across the interface
and picking up a crossing phase $e^{-2\pi i{\bf
a}^TK^{-1}{\bf \lambda}}$ at the interface due to the possible presence of a 
qp $\lambda$ localized at the defect. The ends of the interface mark the presence of a twist defect
from which the branch cuts across which qps transform emanate.

We can see that for our particular interest in the fermion parity flip symmetry there is no clear microscopic representation of an  anyon transformation by this symmetry unless we augment the K-matrix in a stably equivalent way. This leaves many interesting possibilities open for future studies of defects in stably equivalent theories.  
\section{Twist defects}\label{sec:twistdefect}
Now that we have seen how we can represent our fermion parity flip symmetry in the K-matrix formalism (at least for Laughlin states), we are ready to classify the properties of the associated twist defects. 
Twist defects are semiclassical defects which act as static fluxes that permute the
anyon labels as they encircle the defect. 
In contrast to the anyons, they are not dynamical excitations of a quantum Hamiltonian (at least at this stage).
Twist defects are attached to physical branch cuts along which anyon quasiparticles get permuted.
The precise location of the twist defects and branch cuts depend on the microscopic details
of the system in question. For example, in Fig. \ref{fig:SCTOlayer}, the twist defects reside at the 
ends of the trench, and  the gapped trench plays the role of the
branch cut.
 In
other examples,
\cite{Bombin,TeoRoyXiao13long,BarkeshliQi,YouWen}
twist defects lie at the site 
of lattice dislocations and the branch cut is a line of lattice mismatch. Our treatment in this section will primarily focus on the
effective theory without worrying about the microscopic details. A recent
review \cite{teo2015globally} covers many twist defect types and contains more
mathematical details than this present article.

\begin{figure}[htbp]
\centering
\subfigure[ Twist defect $\sigma$ with a branch cut emanating from it. A
passing anyon $\mathbf a$ is mutated to $W \mathbf a$ according to an Anyonic
Symmetry $W$.]{
\includegraphics[width=0.20\textwidth]{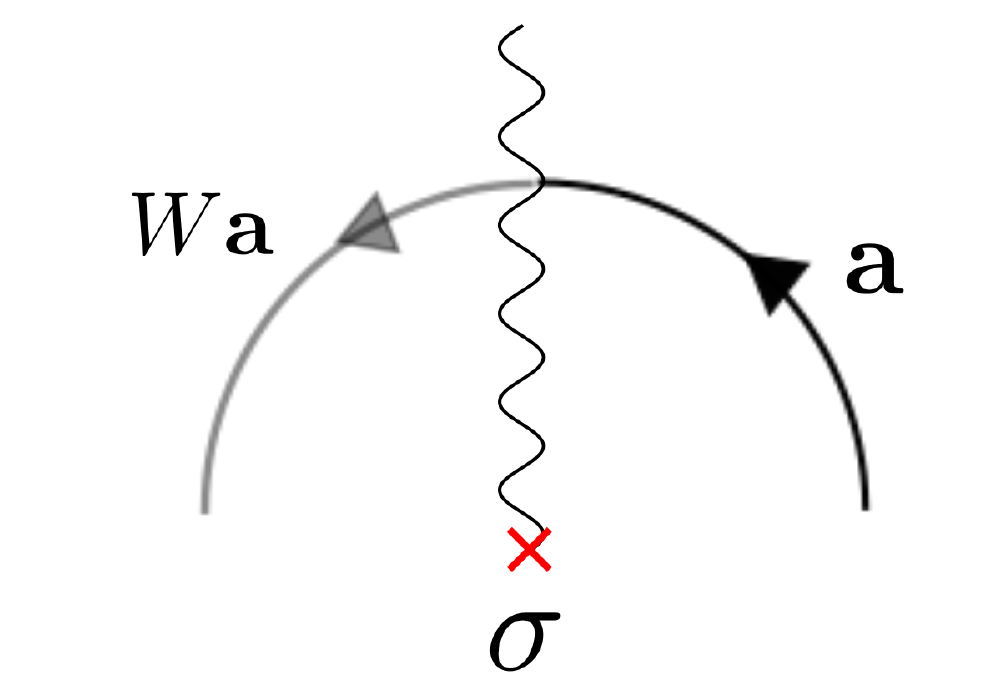}
\label{fig:twistbranch}}
\hfill
\subfigure[\ Fermion Parity Flip symmetry acting on a qp. $\mathbf a$ with vorticity $p$.
$p$ fermions are pumped into it as it crosses the branch cut. ]{
\includegraphics[width=0.20\textwidth]{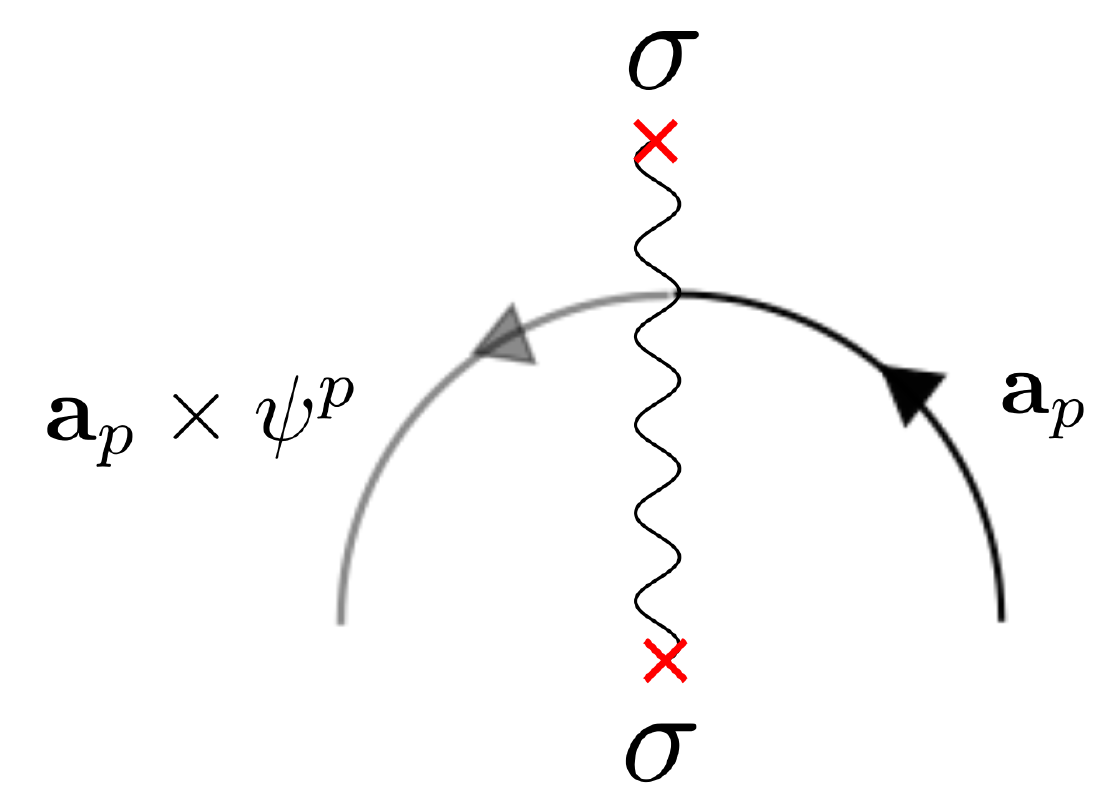}
\label{fig:FermionParityflip}}
\caption{(a) General picture of the action of a twist defect (b) Twist defect for fermion parity flip anyonic symmetry.}\label{fig:twistdefectdef}
\end{figure}

We saw in Sec. \ref{sec:AS} that the action of an AS in $N$-component Abelian systems can be expressed
as an $N\times N$ matrix $W$ acting on the $N$-component vector of topological charges
of a qp  $\mathbf{a}$. (see Fig. \ref{fig:twistbranch}).
In this section we will be working mostly with fermion parity flip symmetry, (c.f., Eq. 
\ref{eqn:fermionparityflipas}) as illustrated  in Fig. \ref{fig:FermionParityflip}.

\begin{figure}[htbp]
\centering
\subfigure[\ Double loop $\Theta_{\m}^\lambda$ measurement of 
the species label associated with the fermion parity flip AS in a Laughlin
state at $\nu=1/(2n+1)$.]{
\includegraphics[scale=0.4]{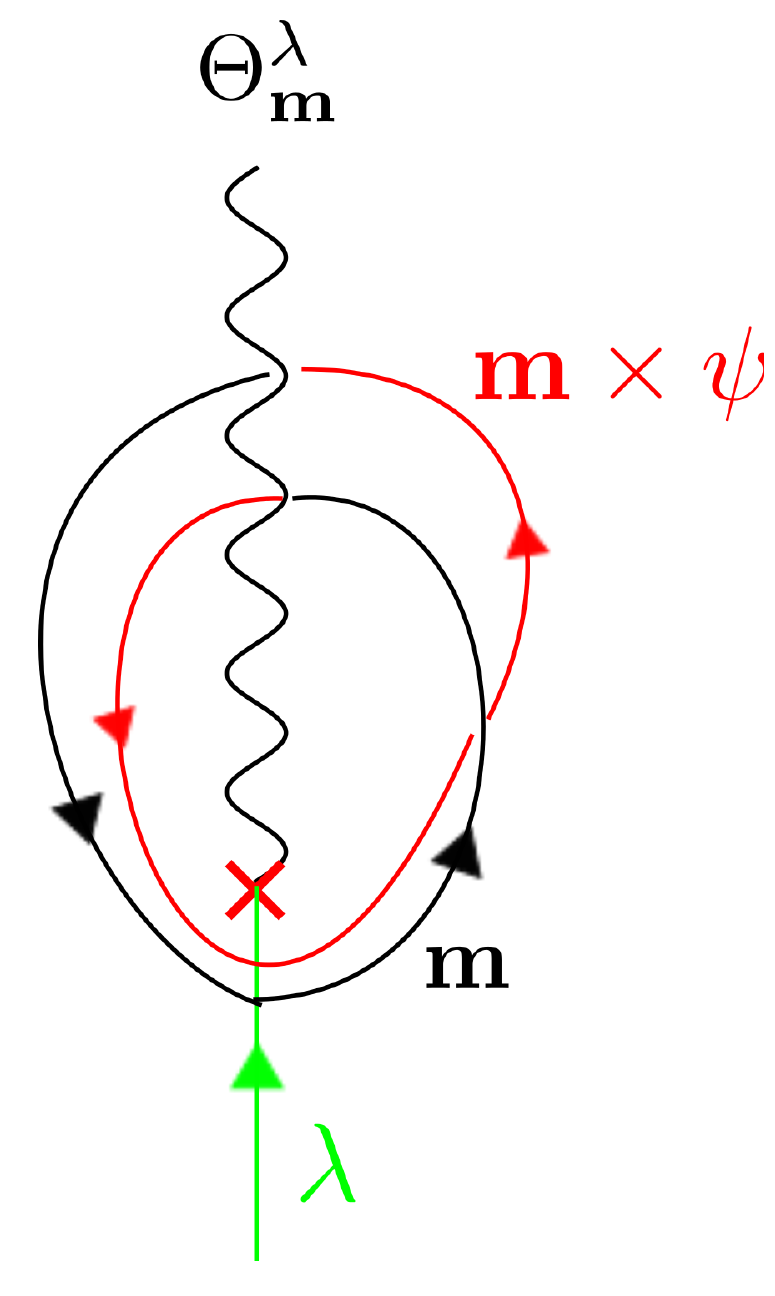}
\label{fig:DoubleLoop}}
\hfill
\subfigure[\ Self-consistency conditions which determine the species
labels associated with an arbitrary AS $W$.]{
\includegraphics[scale=0.27]{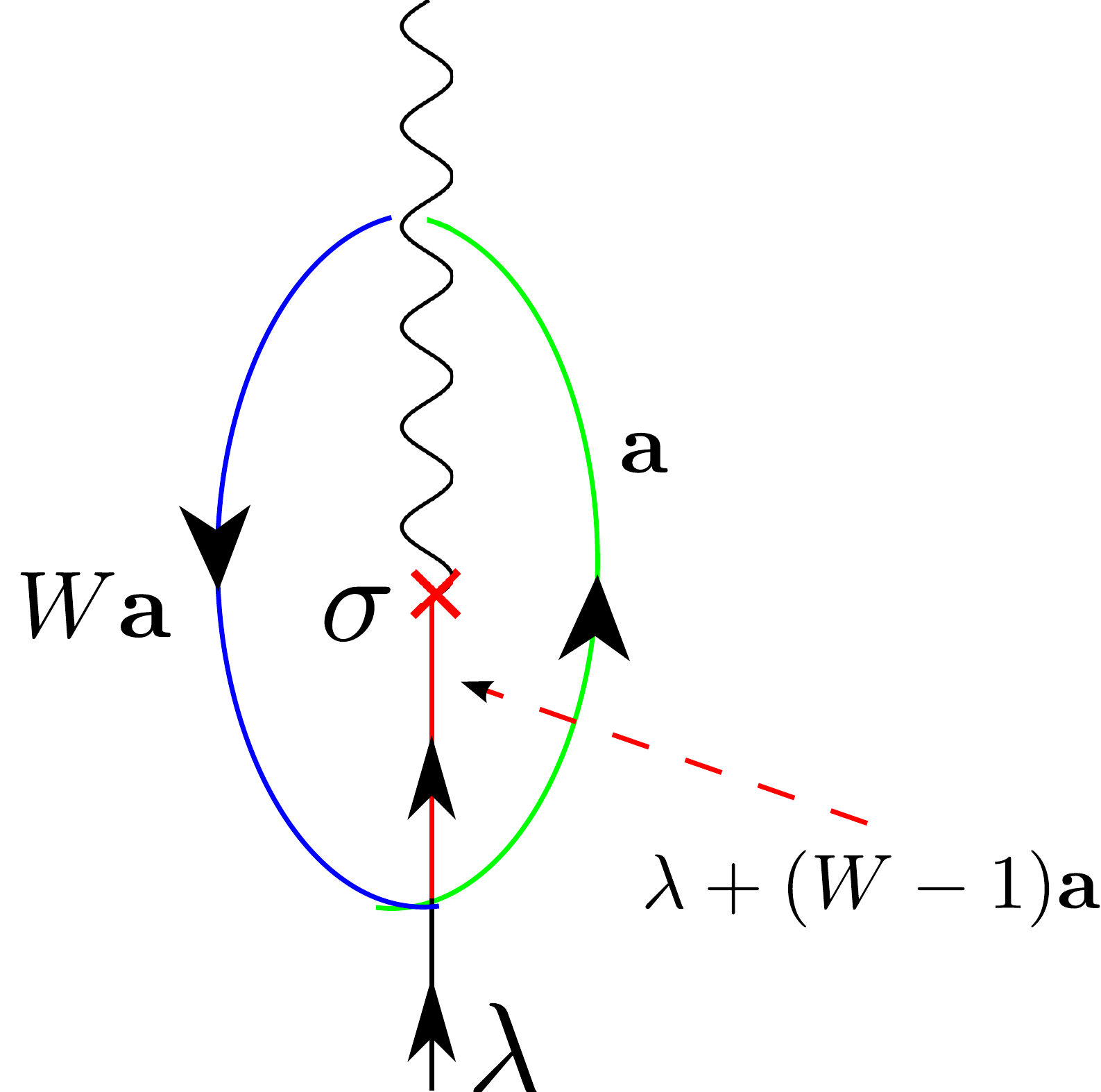}
\label{fig:Defectspecies}}
\caption{Diagrams used for determining defect species.}\label{fig:defectspecies}
\end{figure}

From Eq. \ref{condensate}
we inferred that twist defects can be decorated with anyon labels 
representing a qp localized at the defect. 
These qps can be detected by a braiding measurement, but not all qp attachments give different measurements. The intuition that
defect species labels are synonymous with the full set of qps. Let us consider the $\mathbb{Z}_2$ fermion parity flip
symmetry for a Laughlin state at filling $\frac{1}{2n+1}$. Now, the half-quantum flux $\m$ must be dragged twice around
a defect before it can close in on itself, this corresponds to the double loop $\Theta_{\m}^{\lambda}$ where $\lambda$ 
is the anyon  string attached to the twist defect (see Fig. \ref{fig:DoubleLoop}).
However, it is easy to see that 
\begin{align}
	\Theta_{\m}^{\lambda\m^a}&=\Theta_{\m}^{\lambda}\exp i\left(\theta_{\m,\m^a}+\theta_{\m\psi,\m^a}\right)\nonumber\\
	&=\Theta_{\m}^{\lambda}\exp i\left(2\theta_{\m,\m^a}\right)\left(-1\right)^a\nonumber\\
	&=\Theta_{\m}^{\lambda}\exp\left(\frac{\pi i a}{2n+1}\right)\left(-1\right)^a\nonumber\\
	\implies \Theta_{\m}^{\lambda}&=\Theta_{\m}^{\lambda\m^{2n+1}}
\label{eqn:twistdefectspeciesbraid}
\end{align}
so that a $\lambda$ or $\lambda{\bf m}^{2n+1}$ qp attached to the defect cannot be topologically distinguished by the double loop $\Theta_{\bf m}$ and hence yield identical defect species.

Since the Laughlin qp $\mathcal{E}={\bf m}^2$ is invariant under the AS, it does not mutate after a single cycle around the defect and one can form the single loop $\Omega^\lambda_{\mathcal{E}}$. This obeys \begin{align}\Omega^{\lambda{\bf m}^b}_{\mathcal{E}}&=\Omega^{\lambda}_{\mathcal{E}}e^{i\theta_{{\bf m}^b,{\bf m}^2}}=\Omega^{\lambda}_{\mathcal{E}}\exp\left(\frac{2\pi ib}{4n+2}\right).\end{align} The smallest $b$ that leaves the single loop $\Omega_{\mathcal{E}}$ invariant is $4n+2$, which corresponds to the electron $\psi=\m^{4n+2}$.
The twist defect label $\lambda$ is thus defined up to an electron, $\lambda\psi\equiv\lambda$ in the case of the
fermion parity gauged Laughlin state at filling $\nu=1/(2n+1)$, described by
$K=8n+4$. Thus, we assign species labels $\mu=0,1,\cdots,4n+1$ to the twist defects, each species label identifying
two qp labels with differing fermion parity.
\begin{align}
	\sigma_{\mu}&=\sigma_{\mu}\times\psi\nonumber\\
	\sigma_{\mu}=\sigma_0\times\m^{\mu}&=\sigma_0\times\m^{\mu}\psi
\label{eqn:speciesLaughlin}
\end{align}
where $\sigma_0$ indicates the bare defect.

In any Abelian TO state, the species labels can be worked out for a twist
defect \cite{TeoRoyXiao13long,khan2014}. The nature of a twist defect 
 leads to important constraints that must be satisfied if a qp is to successfully fuse with the defect. This constraint is shown in Fig. \ref{fig:Defectspecies}. Essentially, the twist defect
$\sigma$ transforms an attached qp $\mathbf{a}$ to $W\mathbf{a}$, hence it must have an internal
structure which allows it absorb the difference. Hence, we can write this as
\begin{align}
	\sigma\times\lambda=\sigma\times\left(\lambda+(W-1)\mathbf{a}\right);
	\forall \mathbf{a}\in \mathcal{A};\,\lambda\in\mathcal{A}.
\label{eqn:speciesredundancy}
\end{align}
This interpretation of the twist defect having internal structure provides
a useful way to understand why twist defects have quantum dimension $d>1$. The allowed
defect species $\mu$ are classified by the quotient group
\begin{align}
	\mu\in\frac{\mathcal{A}}{(W-1)\mathcal{A}}
\label{eqn:speciesquotient}
\end{align}\noindent (see Ref. \onlinecite{Teotwistliquids} for more details). 

In the case of the fermion parity flip symmetry we can just calculate this to find
\begin{align}
	(W-1)\mathbf{a_p}=& 1\; \mbox{or}\;\psi;\quad \mbox{for vorticity}\, p\,\,\mbox{even/odd}\nonumber\\
	\mu=&\frac{\mathcal{A}}{\left\{1,\psi\right\}}.
	\label{eqn:fermionparityspecies}
\end{align}
Hence we conclude that the number of defect species is equal to the number of
conjugacy classes $[\mu]$, i.e., 
$|\mathcal{A}|/2$.

To precisely calculate the twist defect
quantum dimensions  we will calculate the increase in ground state
degeneracy associated with a twist defect $\sigma$ and hence infer its quantum
dimension $d_\sigma$ using methods borrowed from Refs. \onlinecite{TeoRoyXiao13long,BarkeshliJianQi13long}.
\begin{figure}[htbp] \begin{center}
\includegraphics[width=0.4\textwidth]{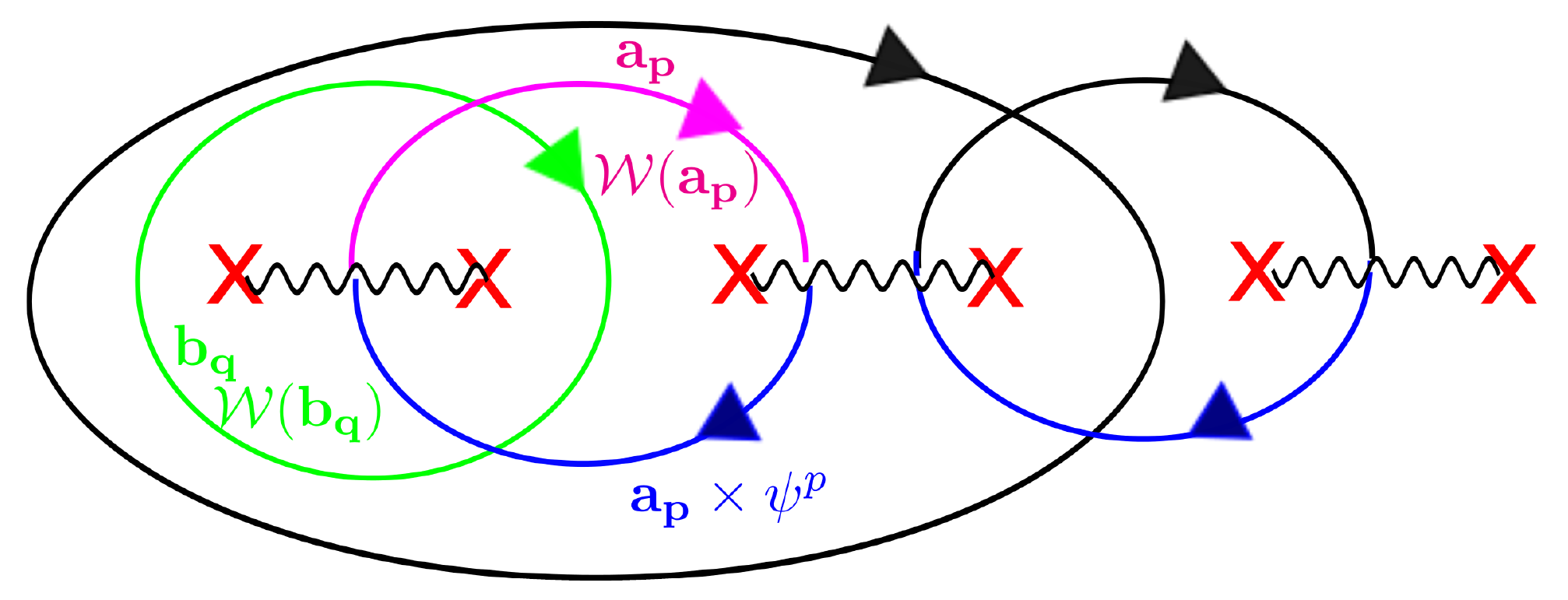} \end{center}
\caption{(color online) Twist
defect configuration and  Wilson loops $\mathcal{W}$ corresponding to fermion
parity flip. The algebra generated by the Wilson loops determines the ground
state degeneracy in the presence of twist defects.} \label{fig:Wilsonloops}
\end{figure}
For example, for an MBS the expected value of $d_{\sigma}$ is $\sqrt{2}$, and we hope to recover this from the algebra of Wilson-loop observables.
In particular, consider the twist defect configuration in Fig.
\ref{fig:Wilsonloops}. There are in general $n$ independent branch cuts and
$2n$ twist defects at which the branch cuts terminate. We consider the Wilson
line algebra of $\mathcal{W}(\mathbf{a_p})$ and $\mathcal{W}(\mathbf{b_q})$,
where the subscripts $p$ and $q$ denote the vorticities of $\mathbf{a}$ and
$\mathbf{b}$ respectively

\begin{align}
	\mathcal{W}(\mathbf{a}_p)\mathcal{W}(\mathbf{b}_q)&=\mathcal{W}(\mathbf{b}_q)\mathcal{W}(\mathbf{a}_p)\nonumber\\
	&\exp
	i\left[\theta(\mathbf{a}_p,\mathbf{b}_q)-\theta(\mathbf{a}_p,\mathbf{b}_q)+\theta(\psi^p,\mathbf{b}_q)\right]\nonumber\\
	&=\mathcal{W}(\mathbf{b}_q)\mathcal{W}(\mathbf{a}_p)(-1)^{pq}
\label{eqn:Wilsonalgebra}
\end{align} where  $\theta(\mathbf{a}_p,\mathbf{b}_q)$ is the
braiding phase of the qps $\mathbf{a}_p$ and $\mathbf{b}_q$. 
Thus, if both $p$ and $q$ are odd
$\mathcal{W}(\mathbf{a}_{\mbox{odd}})\mathcal{W}(\mathbf{b}_{\mbox{odd}})=(-1)\mathcal{W}(\mathbf{b}_{\mbox{odd}})\mathcal{W}(\mathbf{a}_{\mbox{odd}}).$

Since the Wilson lines map the ground state (manifold) onto itself, the states that span
the ground state manifold must form an irreducible representation of the Wilson
line algebra.
The smallest representation of the above algebra is $2$ dimensional
\begin{align}
	\mathcal{W}(\mathbf{a}_{\mbox{odd}})\ket{i}&=(-1)^{i}\ket{i},\,\,i\in[0,1]\nonumber\\
	\mathcal{W}(\mathbf{b}_{\mbox{odd}})\ket{i}&=\ket{i+1;\mbox{mod }2}.
\label{eqn:2dimWilsonalgebra}
\end{align}
The algebra generated by $\mathcal{W}(\mathbf{a}_{\mbox{odd}})$ is essentially
generated by $\m$. This follows because vorticity is additive under fusion
and hence $\forall \,\,\mathcal{W}(\mathbf{a}_{\mbox{odd}})\,\,\exists
\,\,\mathcal{W}(\mathbf{a}_{\mbox{even}})\,:\,\mathcal{W}(\mathbf{a}_{\mbox{odd}})=\mathcal{W}(\mathbf{a}_{\mbox{even}}\times \m)$.
For $2n$ such twist
defects on a closed sphere there are $n-1$ copies of the Wilson line algebra. Hence, as $n\
\rightarrow\infty$ we get the quantum dimension $d_{\sigma}=\sqrt{2}$.

\begin{figure}[htbp] \begin{center}
\includegraphics[width=0.5\textwidth]{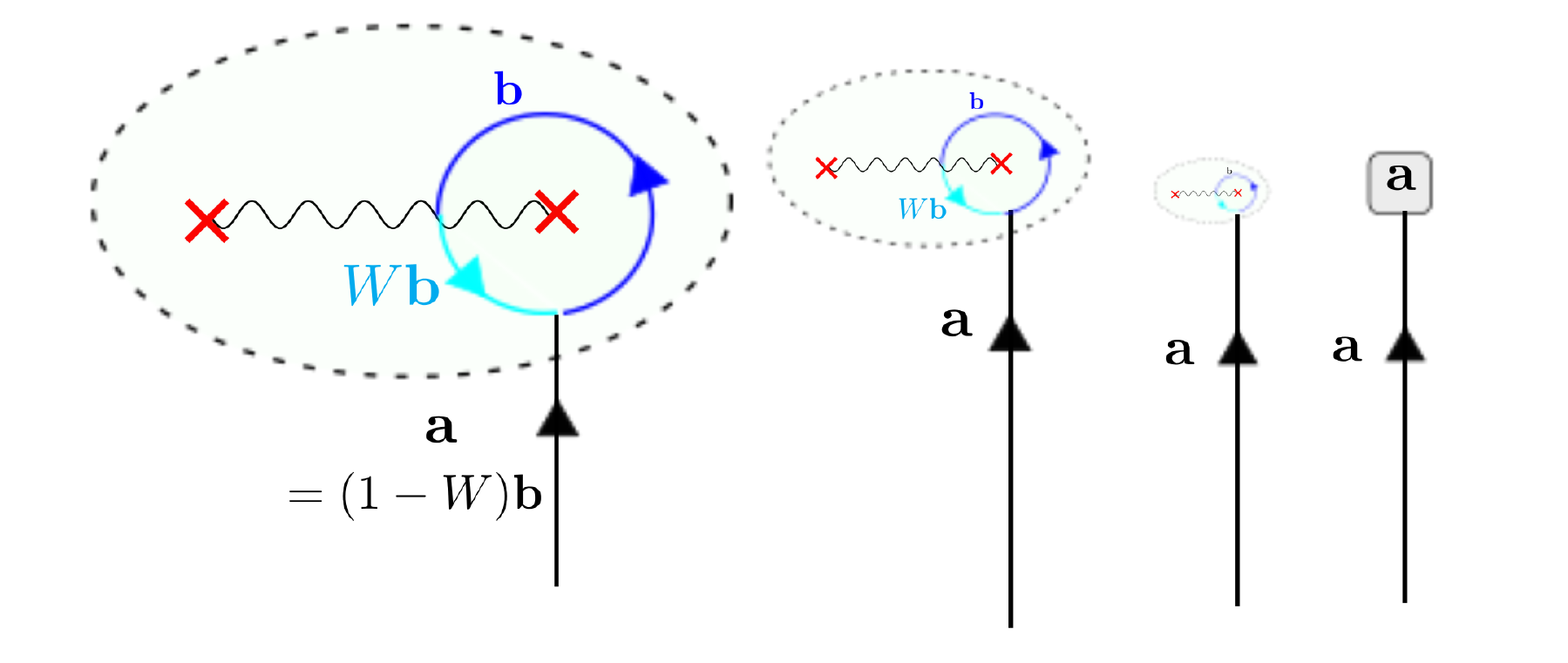} \end{center}
\caption{(color online)Fusion of conjugate defects leading to Abelian fusion
channels. As we go farther away from the twist defects ignoring the
microscopic details, we just see the qp. strings terminating at the site of
defect fusion.} \label{fig:fusion}
\hspace{-1pt}
\end{figure}

A quantum dimension larger than $1$ is generally associated with multichannel fusion. To show this let us 
consider Fig. \ref{fig:fusion}.
which involves the fusion of a defect associated
with an anyonic symmetry operation $W$ with its conjugate defect $W^{-1}$, so that the branch cuts can cancel each other. Since fermion
parity flip is a $\mathbb{Z}_2$ symmetry, the inverse defect is the same as the original defect.
The fusion outcome of a pair of conjugate defects must be a trivial defect in the sense that there
is no remaining qp permutation. However, the overall
fusion outcome depends upon how qp strings are attached between the defects, in effect, taking
into account the defect species. Fig. \ref{fig:fusion} shows how the overall open
string contributes $(1-W)\mathcal{A}$ to the defect fusion pair. As we zoom
further and further away from the defect pair and gradually ignore the details
of the action of the defects on the encircling qps it seems as if the $\mathbf{a}$
string comes and terminates at the defect. Thus, this represents an Abelian fusion
channel outcome.


To determine the possible fusion
outcomes when we fuse conjugate defects, we need to determine the 
distinct quasiparticle strings that can be hung between the defect pair. For
example, consider
Fig. \ref{fig:fusion} with the fermion parity flip AS. If the quasiparticle $\bb{b}$
has odd/even vorticity, the outcome string $\bb{a}=(1-W)\bb{b}$ is
$\psi$ or $1(\text{vacuum})$. This determines the outcomes for bare defects. If there are additional
defect species, their fusion differs by the corresponding attached qps that mutated the defect species from the bare defect species. These attached qps need to be fused into the outcome for the bare defect
to find the result for non-trivial defect species. This is summarized in the following fusion rules:
\begin{align}
	\sigma_0\times\sigma_0&=1+\psi\nonumber\\
	\sigma_\mu\times\sigma_\nu&=\mu\times\nu\times(1+\psi).
	\label{eqn:fusiondefectfermionparityflip}
\end{align}
Thus, we see that the fusion is the same as that of non-Abelian Ising defects, and the 
overall fusion always results in two possible outcomes with opposite fermion parities.
Also, as a useful check to see if all the possible fusion channels
have been accounted for, we note that quantum dimension on both sides of Eq. 
\ref{eqn:fusiondefectfermionparityflip} matches up, i.e.,
$\sqrt{2}\times\sqrt{2}=1+1$.
As a check, we note that \eqref{eqn:fusiondefectfermionparityflip} is independent from the ambiguity in the choices of defect labels $\mu\equiv\mu\times\psi$, $\nu\equiv\nu\times\psi$ because $\psi\times(1+\psi)=1+\psi$, i.e., independent of which anyon we choose to act as a species label.

\subsection{Charge Conjugation Defect}\label{subsec:chargeconjugationdefect}
For completeness, we also reproduce some of the results for charge conjugation
defects that were presented in Refs. 
\onlinecite{ClarkeAliceaKirill,LindnerBergRefaelStern,MChen,Vaezi} in the context of
a Laughlin state in proximity so an s-SC. The resulting state after fermion parity gauging is $K=8n+4$, and we can work out the details of the twist defects of the ungauged theory from this more exotic emergent phase.
A charge conjugation twist defect acts as $\m\rightarrow -\m$.
Using Eq. \ref{eqn:speciesredundancy}, we find for the defect species the rule
\begin{align}
\sigma_\mu&=\sigma_\mu\times\m^2\nonumber\\
\label{eqn:conjugationdefectspecies}
\end{align} which implies that there are only two defect
species, $\sigma_0$ and $\sigma_1$,
\begin{align*}
\sigma_0=\sigma_0\times\m^2&=\sigma_0\times\m^4=\sigma_0\times\m^6\cdots\nonumber\\
\sigma_1=\sigma_0\times\m=&\sigma_0\times\m^3=\sigma_0\times\m^5\cdots
\end{align*}
These twist defects have quantum dimension $\sqrt{4n+2}$ and give rise to fusion
rules
\begin{align}
\sigma_0\times\sigma_0=\sigma_1\times\sigma_1&=1+\m^2+\m^4\cdots\nonumber\\	
\sigma_0\times\sigma_1&=\m+\m^3+\m^5+\cdots
\label{eqn:fusionchargeconjugationdefects}
\end{align} We will use these results at the end of the next section when we discuss gauging the charge-conjugation AS. 

\subsection{Experimental Consequences}
From this analysis we arrive at another main result of our work, i.e., the two general AS available in fermionic FQH states give rise to very different twist defect NAZMs. For fermion parity flip AS we \emph{always} find MBS, and for charge conjugation, the NAZM depends on the parent state. For the Laughlin states we recover the family of parafermion NAZMs with quantum dimension $d_{\sigma}=\sqrt{4n+2}=\sqrt{2/\nu}.$ In Laughlin state-s-SC heterostructures it is possible to find both types of defects, and which defects are stabilized depends on the local interactions near the defect. As a proof of principle, we show in Appendix \ref{app:relevance} that it is possible to tune the interactions for the $\nu=1/3$ Laughlin state such that MBS are stabilized instead of a $\mathbb{Z}_6$ parafermion. Hence, it will be interesting to see which NAZM appears in possible experimental realizations.

\section{Gauging the anyonic Symmetry}\label{sec:Asgauge}
We will now move on to investigate something more theoretical in nature that is a natural extension of our work so far. 
Until now we have treated an AS as a \emph{global}, discrete symmetry of the qp set $\mathcal{A}$ of a given
topological phase. The twist defects that we have identified are semiclassical, static fluxes that were put into the system by hand.  Gauging the AS promotes the twist defects to deconfined quantum excitations of a new topological phase, which is non-Abelian. We now move on to identifying the resulting set of topological qps after gauging
the AS. We will denote this phase as  a twist liquid, and such phases have been treated
comprehensively recently in Refs. \onlinecite{barkeshli2014,Teotwistliquids}. A 
treatment of the full structure of twist liquids that can emerge from fermionic FQHE states is beyond our scope. We have narrowed our focus on just gauging the fermion parity flip symmetry or the usual charge-conjugation symmetry, and furthermore, we only enumerate the emergent qp structure without deriving the full braiding matrices and F-matrices of the resultant braided fusion category. In this section we will review the algorithm for gauging a conventional discrete symmetry and an AS, and then we will enumerate the twist liquid qps for gauging the fermion parity flip symmetry and the charge-conjugation symmetry for the Laughlin series.  

Let us begin with a discussion of gauging a discrete symmetry with a trivial background vacuum. 
We will denote  a discrete gauge theory in 2+1 dimensions based on the group $G$  as
$D(G)$; it is  also called the quantum double of $G$\cite{Preskilllecturenotes,Bais-1992,Freedman2004,propitius95,propitius1995topological,mochon2004}. The excitations are called anyons and
are denoted by the 2-tuple
\begin{align*}
\chi=\left([W],\rho\right).	
\end{align*}
The flux component $[W]$ is given by a conjugacy class $[W]$ of an element $W$ in $G$ defined as
\begin{align*}
	[W]=\{W':W'=NWN^{-1},N\in G\}.
\end{align*}
Given the flux $[W]$, the second component gives an associated value of the  charge which is labeled by the
irreducible representations $\rho$ of the centralizer $C_{G}([W])$
\begin{align*}
	C_G([W])=\{N\in G:NW=WN\}.	
\end{align*}
Note that the definition of the centralizer is independent  (up to adjoint isomorphism) of which representative
 of the conjugacy class $[W]$ is chosen.
The allowed charged components are thus characterized by an irreducible
representation $\rho:C_G([W])\rightarrow U(\mathcal{N}_\rho)$. Enumerating all possible combinations of $[W]$ and $\rho$ yields the full set of flux, charge, and dyonic qps of the deconfined phase of the discrete $G$ gauge theory. 

Now, consider a more complicated scenario where we have a parent Abelian topological state with an initial quasiparticle set
$\mathcal{A}$ and a discrete, global AS group $G$. The anyon excitations of the $G$ twist
liquid obtained by gauging the global AS group $G$ are composites of the
flux and charge of the gauged AS $G$ and qps of the parent Abelian theory; but there are requisite 
constraints that must be satisfied by the composites. The twist liquid has anyons
 $\chi$ which can be labeled using  a 3-tuple
\begin{align*}
	\chi=([W],\pmb{\lambda},\rho).	
\end{align*}
The flux label $[W]$ is specifies  a conjugacy class of $G$. Given $W$, a
representative element of the conjugacy class $[W]$, we form the set of defect
species $\mathcal{A}_W=\frac{\mathcal{A}}{(W-1)\mathcal{A}}$.
$\pmb{\lambda}$ is an \emph{orbit} of the anyon labels $\lambda$ drawn
from $\mathcal{A}_W$. Precisely, $\pmb{\lambda}$ is the $C_{G}([W])$ orbit in
$\mathcal{A}_W$.
Thus
\begin{align}
	\pmb{\lambda}=\lambda_1+\lambda_2+\lambda_3+\cdots+\lambda_l,\lambda_i\in\mathcal{A}_W.
	\label{eqn:supersector}
\end{align}
Thus, the elements in $C_{G}([W])$ can permute the elements $\lambda_i$
while keeping $\pmb{\lambda}$ unchanged.

The charge component $\rho$ is characterized by an $\mathcal{N}_\rho$
dimensional, irreducible representation of a \emph{restricted} centralizer of the flux
$W$ and orbit $\pmb{\lambda}$ $C_G^{\pmb\lambda}([W])$.
\begin{align}
	C_G^{\pmb\lambda}([W])&=\{N\in C_G([W]):N\lambda_1=\lambda_1\}.
\label{eqn:restrictedcentralizer}
\end{align}
One might worry that Eqs. \ref{eqn:supersector} and \ref{eqn:restrictedcentralizer}
are dependent on the choice of species label $\lambda_1$, but different choices
are related by adjoint isomorphisms and lead to identical anyon content\cite{Teotwistliquids}.

While these mathematical definitions might seem a bit opaque, they are simple to apply for our cases of interest. Let us now use the above formalism to determine the qp content of the gauged  fermion-parity flip or
charge conjugation theories when starting from a Laughlin state at $\nu=1/(2n+1)$ with gauged fermion parity, i.e., the theory ultimately described by $K=8n+4$ as described in the main text.

\subsection{Fermion parity flip}
 The fermion parity flip acts on the qp set as
$\m^p\rightarrow\m^p\times\psi^p$. 
As this is a $\mathbb{Z}_2$ symmetry,  the AS group to be gauged is
$G=\mathbb{Z}_2$. This group has  two conjugacy classes: $[1]$ (the trivial conjugacy class) and $[\sigma]$
(the non-trivial flux/twist defect around which anyon labels get permuted).

For the trivial flux, $[1]$, the species labels are taken as orbits from the set
$\mathcal{A}_{[1]}=\mathcal{A}=\{1,\m,\m^2,\cdots,\m^{8n+3}\}$.  The appropriate $\mathbb{Z}_2$ orbits of this set of qps can be summarized as 
\begin{align}
	\mathbb{Z}_2 \m^{2a}&=\m^{2a};\,\, a\in [0,1,\cdots,4n+1]\nonumber\\
	\mathbb{Z}_2
	(\m^{2a+1}+\m^{2a+1}\psi)&=(\m^{2a+1}+\m^{2a+1}\psi);a\in[0,\cdots,2n]
\label{eqn:orbitsfpf}
\end{align} That is, even powers of $\m$ form an orbit by themselves, and odd
powers form an orbit with itself fused with a fermion.
For an Abelian group the centralizer of each element is the whole group so the choice of flux does not restrict the possible charge representations. However, the choice of species does restrict the charges and we have the restricted centralizer subgroups
\begin{align*}
	C_{\mathbb{Z}_2}^{\m^{2a}}([1])&=\mathbb{Z}_2\nonumber\\
	C_{\mathbb{Z}_2}^{(\m^{2a+1}+\m^{2a+1}\psi)}([1])&=\mathbb{Z}_1
\end{align*}\noindent where by $\mathbb{Z}_1$ we mean the group containing only the identity. The group $\mathbb{Z}_2$   has two representations $\rho_{+}$ and $\rho_{-}$ which can be held by qps which are even powers of $\m,$ whereas odd powers of $\m$ (that form orbits) cannot hold any non-trivial charge. 

For the non-trivial flux sector $[\sigma]$ there are $4n+2$ species labels 
$\lambda_i=0,1,2,\cdots,4n+1$, where $\lambda_i$ can imply either $\m^{i}$ or $\m^i\times\psi$.
Each  $\lambda_i$ is invariant under the $\mathbb{Z}_2$ action and 
forms a $\mathbb{Z}_2$ orbit by itself. As such, the restricted centralizer group is
$C^{i}_{\mathbb{Z}_2}([\sigma])=\mathbb{Z}_2$.

The full set of qps,  along with their quantum dimensions are
\begin{align}
&\left([1],\m^{2a},\rho_{+}\right); d=1\nonumber\\
&\left([1],\m^{2a},\rho_{-}\right); d=1\nonumber\\	
&\left([1],\m^{2a+1}+\m^{2a+1}\psi,\rho_+\right); d=2\nonumber\\	
&\left([\sigma],\lambda_i,\rho_{+}\right); d=\sqrt{2}\nonumber\\	
&\left([\sigma],\lambda_i,\rho_-\right); d=\sqrt{2}.	
\label{eqn:qplabels}
\end{align}
Adding up the total quantum dimension we find
$D=\sqrt{\sum_{\chi}d_{\chi}^2}=2\sqrt{8n+4}$, which is exactly what we expect
when a 2 fold symmetry is gauged in a topological phase with initial quantum dimension
$\sqrt{8n+4}$. 

This theory is non-Abelian as well since some qps have $d>1.$ 
 It can be identified with the tensor product theory \begin{gather}SO(N_1)_1\otimes SO(N_2)_1\otimes\mathbb{Z}_{2n+1}^{(2)}\\\mbox{for $N_1,N_2$ odd and $N_1+N_2=4n+2.$}\nonumber\end{gather} Here the $SO(N)_1$ is an Ising-like state with chiral central charge $c_-(SO(N)_1)=N/2$. It contains anyons $\{1,\psi,\sigma\}$, where $\psi$ is identified with one of the fermions $([1],{\bf m}^{4n+2},\rho_\pm),$ and $\sigma$ is identified with either $([\sigma],\lambda_0,\rho_+)$ or $([\sigma],\lambda_{2n+1},\rho_+)$. They have spins $h_\psi=-1/2$ and $h_\sigma=N/16$ respectively, and follow the fusion rules $\sigma\times\psi=\psi$ and $\sigma\times\sigma=1+\psi$. The $\mathbb{Z}_{2n+1}^{(2)}$ state\cite{MooreSeiberg89,Bondersonthesis} is Abelian with chiral central charge $c_-(\mathbb{Z}_{2n+1}^{(2)})=-2n$. It has qps $\{1=E^0,E^1,\ldots,E^{2n}\}$, where $E^j$ is identified with $([1],{\bf m}^{4j},\rho_+)$. They have spins $h_{E^j}=2j^2/(2n+1)$ and follow the fusion rules $E^i\times E^j=E^{i+j\;\mathrm{mod}\;2n+1}$.

Let us compare this result to what we would find by gauging charge-conjugation symmetry.

\subsection{Charge conjugation symmetry}
The charge conjugation symmetry acts as $\m^{a}\rightarrow\m^{-a}$.
Again, this discrete symmetry group is $\mathbb{Z}_2$.
For the trivial flux sector $[1]$ 
 the orbits constructed from the full anyon set $\mathcal{A}$ are
\begin{align}
	\left\{{ 1},\{\m^a,\m^{-a}\}, {\psi}\right\}; \,\, a\in[1,2,\cdots, 4n+1]	
\label{eqn:orbitschargeconjugation}
\end{align}
The restricted centralizer groups are
\begin{align*}
	C_{\mathbb{Z}_2}^{\{1\}}([1])&=C_{\mathbb{Z}_2}^{\{\psi\}}([1])=\mathbb{Z}_2\nonumber\\
	C_{\mathbb{Z}_2}^{\{\m^a+\m^{-a}\}}([1])&=\mathbb{Z}_1;\,\, a \in[1,2,\cdots,4n+1].
\end{align*} Hence the lone qp orbits can hold either charge, while the two-particle orbits can only hold trivial charge. 

For the flux component $[\sigma],$ the relevant species labels are $A_{\sigma}=\frac{\mathcal{A}}{(\sigma-1)\mathcal{A}}=\{0,1\}$. 
The species $0$ can represent any qp $\m^{2a}$ while $1$ could be any qp labelled
by $\m^{2a+1}$ (c.f., Sec. \ref{subsec:chargeconjugationdefect}).
The restricted centralizer groups are
\begin{align*}
	C_{\mathbb{Z}_2}^{\{0/1\}}([\sigma])=\mathbb{Z}_2. 	
\end{align*}

Hence, the full set of twist liquid qps are 
\begin{align}
&\left([1],1,\rho_{+}\right);\,\,\left([1],1,\rho_{-}\right); d=1\nonumber\\
&\left([1],\psi,\rho_{+}\right);\,\,\left([1],\psi,\rho_{-}\right); d=1\nonumber\\
&\left([1],\m^a+\m^{-a},\rho^{+}\right); a=1,2,\cdots,4n+1; d=2\nonumber\\
&\left([\sigma],i,\rho_+\right); \left([\sigma],i,\rho_-\right); i\in[0,1];
d=\sqrt{4n+2}.
\label{eqn:qplabelschargeconjugation}
\end{align}
This agrees with the conformal field theory content of the
$U(1)_{4n+2}/\mathbb{Z}_2$ orbifold.\cite{GINSPARG1988,DijkgraafVafaVerlindeVerlinde99}
Again, the net quantum dimension is $2\sqrt{8n+4}$ because charge conjugation
is also a $\mathbb{Z}_2$ symmetry. However, the net anyon content of the twist
liquid is very different as we now see non-Abelian objects with much higher quantum dimensions.

\section{Conclusions}
 In this paper we have provided a framework from which the exotic
non-Abelian zero modes at domain walls in superconductor-topological phase
heterostructures can be understood. The main idea is the consideration 
 of an extended TO due to coupling of the original state
with a (somewhat artificial) $\mathbb{Z}_2$ gauge theory from the s-wave superconductor. This new
theory reveals the presence of an extra, general anyonic symmetry, the fermion parity
flip symmetry which had been overlooked beforehand. The fermion parity flip
symmetry reveals itself in the form of MBS at corresponding twist
defects. Remarkably, we can extract  hidden information
about the structures of defects in the experimentally accessible un-gauged theory by appealing to the fermion
parity gauged version. As already remarked, this is similar to the
recent trend of investigating ungauged SPT states by looking at their gauged
versions\cite{LevinGubraiding,Braiding3DLevinWang,LevinWang3Dbraiding}.
Finally, we determined
the structure of two families of exotic non-Abelian twist liquids obtained by deconfining the
twist defects and promoting them   
to genuine quantum excitations of the system.

These results shed new light on predictions of parafermions in superconductor/FQH heterostructures since  we have shown that experimental geometries where parafermions
can exist might harbor MBS instead. The outcome depends on the details of the interactions at the
interface, and we have explicitly constructed cases where one non-Abelian object or the other can be stabilized.
Some details about the s-wave proximity effect have been mostly ignored since our treatment has been at the
level of gauge fields. However, gapping out an edge and realizing the fermion
parity flip symmetry as in Eq. \ref{edgecoupling} requires inducing a proximity effect in
oppositely propagating edges at the trench using an s-wave superconductor. Hence, this 
requires them
to have opposite spin polarization and it is unlikely that such TO will
arise in the case of spin-polarized Laughlin states. Such twist defects should
be observable in fractional Chern insulators where counter-propagating
edges have opposite spin polarization, or in FQH 2DEGs with carefully engineered g-factors. We optimistically note that the fermion
parity flip symmetry is
viable in all the geometries proposed to date  in superconducting 
heterostructures where parafermions have been predicted to exist due to charge conjugation
symmetry.\cite{ClarkeAliceaKirill,LindnerBergRefaelStern,MChen,Vaezi}
The remaining open question is then a determination of
experimentally viable methods of tuning forward scattering terms so that
different non-Abelian modes are favored over each other.  
\acknowledgments{We would like to thank V. Chua for discussions.  We thank the National Science Foundation under the grants DMR 1351895-CAR (TLH) and DMR 0644022-CAR (SV) for support.}
\appendix 

\section{K-Matrix theory}{\label{app:Kmatrix}}
The K-matrix formalism  \cite{WenZee92,Wenedgereview} provides a concise way of
describing the effective field theory of any Abelian TO state using
2+1 D Chern Simons (CS) theory.  The bulk action takes the form \begin{align}
\mathcal{L}_{\text{bulk}}&=\frac{K_{IJ}}{4\pi}\epsilon^{\mu\nu\lambda}\alpha_{I\mu}\partial_{\nu}\alpha
_{J\lambda}-\frac{1}{2\pi}t_{I}\epsilon^{\lambda\mu\nu}A_{\lambda}\partial_{\mu}\alpha_{I\nu},\nonumber\\
&\quad I,J=1,2,\cdots,N.  \label{eq:bulkCSconv} \end{align} Here we
have assumed that there are $N$ $U(1)$ gauge fields $\alpha_I$, thus
the $K$ matrix is $N\times N,$ and the gauge group is $U(1)^N$. The $K$
matrix is symmetric and has integral entries. The charge vector
$\bb{t}\in\mathbb{Z}^N$ details how the gauge fields $\alpha$ couple to
the external electromagnetic field $A$.

In this formalism the quasiparticle excitations carry integral gauge charge
$l_I$ under the gauge fields $\alpha_I$. Thus, we will describe a quasiparticle
by an integral vector $\bb{l}$ in the lattice $\Gamma=\mathbb{Z}^N$.
The number of topologically distinct quasiparticles in the theory is the same as the ground
state degeneracy on a torus and is $|\text{det}K|$ (since the theory is Abelian). 
The local
particles in the theory braid trivially around all
other excitations, and belong to the lattice
$\Gamma^*=K\mathbb{Z}^N$.
The fusion of two quasiparticles in this theory amounts to adding lattice
vectors. Thus, $\bb{a\times b=a+b}$.
Quasiparticles (qps) are considered to be equivalent
if they differ only by the addition/fusion of local quasiparticles: \begin{align} \bb{a}\equiv
\bb{a}+K\mathbb{Z}^N. \label{eqn:Kmatrixequivalenceclasee} \end{align}
We denote an equivalence class of qps by the symbol $[\mathbf{a}]$.

Thus, distinct qps are just the different equivalence classes
$\Gamma/\Gamma^*=\frac{\mathbb{Z}^N}{K\mathbb{Z}^N}$.
Equivalent qps have the same topological characteristics (exchange
and braiding statistics), but  might not share some non-topological characteristics
(such as charge, which exists due to symmetry not topology). 

The topological character of the theory is specified by the set $\mathcal{A}$ of distinct qps,
and their braiding and exchange properties, which are contained in the modular $S$ and $T$
matrices. By definition $S_{\bb{a,b}}=\frac{1}{\sqrt{|\mathcal{A}|}}e^{i\theta_{\bb{a,b}}}$, where
$\theta_{\bb{a,b}}$ is the braiding phase of
$\bb{a}$ around $\bb{b}$, and  $T$  is specified by the exchange phase $\delta$ of the
qps:
$T_{\bb{a,b}}=e^{i\delta_{\bb{a}}}\delta_{\bb{a,b}}=e^{2\pi i
h_{\bb{a}}}\delta_{\bb{a,b}}$, where $h_{\bb{a}}$ is the topological spin of the qp
$\bb{a}$ (and which corresponds to the scaling dimension of the corresponding
primary field in the edge conformal field theory).

The electromagnetic
charge of the quasiparticle $\bf{l}$ is \begin{align}
q_{\bb{l}}&=\bb{l}^TK^{-1}\bb{t}\quad(\text{in units of $e$}).
\label{eqn:chargeqpKmatrix} \end{align} 
The filling fraction takes the form \begin{align} \nu=\bb{t}^TK^{-1}\bb{t}.
\label{eqn:fillingKmatrix} \end{align} The expression for the braiding
angle $\theta_{\bb{l,m}}$ is \begin{align}
\theta_{\bb{l,m}}&=2\pi\bb{l}^TK^{-1}\bb{m}
\label{eqn:braidingKmatrix} \end{align} while the exchange
phase  of the quasiparticle $\bb{l}$ is \begin{align}
\delta_{\bb{l}}&=\pi\bb{l}^TK^{-1}\bb{l}.
\label{eqn:exchangeKmatrix} \end{align} 

We note that there are different $K$ matrices which encode the
same information. These different $K$ matrices are defined by a $GL(N,Z)$ transformations of the original $K$ matrix.  Thus,
\begin{align} K\rightarrow WKW^T; t\rightarrow Wt \quad W\in
GL(N,\mathbb{Z}),\,\,|\text{det}W|=1	\label{eqn:equivKmatrix}
\end{align} leaves the physical content of the theory in Eq. \eqref{eq:bulkCSconv} invariant and is simply a
basis change implemented by redefining the $U(1)$ gauge fields
$\alpha\rightarrow W^T\alpha$.

Finally, a natural edge theory corresponding to
Eq. \eqref{eq:bulkCSconv} has Lagrangian density \begin{align}
\mathcal{L}_{\text{edge}}&=\frac{K_{IJ}}{4\pi}\partial_x\phi_I\partial
_t\phi_J+\frac{e}{2\pi}\epsilon^{\mu\nu}t_I\partial_{\mu}\phi_IA_\nu-\nonumber\\&
V_{IJ}\partial_x\phi_I\partial _x\phi_J.\label{eq:edgeCS} \end{align}
Here, $\partial_\mu\phi_I=\alpha_{I\mu }$\cite{Wenedgereview} and
$V_{IJ}$ is a non-universal positive-definite ``velocity" matrix which
encodes forward-scattering interactions between the edges.  Corresponding to every
anyon $\bb{a}$ in the bulk there is a vertex operator
$e^{i\bb{a}.\phi}$ in the edge conformal field theory.
 \section{Gauging Fermion Parity in Laughlin States Coupled to Superconductors, an approach inspired by stable equivalence}
\label{sec:Laughlinstabequiv} 
In this section we provide an alternative interpretation of the emergence of
$K_{\text{sc}}=8n+4$ state from a Laughlin state at filling $\frac{1}{2n+1}$ using
ideas of stable equivalence.\cite{cano2013bulk}.

From Eq. \ref{eqn:eqn1}, we have the relevant
action for a  Laughlin state in contact with
a topological s-wave superconductor:
\begin{align} \mathcal{L}=\frac{2n+1}{4\pi}\alpha \wedge d\alpha
-\frac{1}{2\pi}a\wedge d\alpha+\frac{1}{\pi}a\wedge db.
\label{eqn:eqnLaugh} \end{align}
We can rewrite the above Lagrangian in a
$K$-matrix form using the full basis $\beta=(\alpha,a,b)^T$: \begin{align}
\mathcal{L}&=\frac{1}{4\pi}K_{IJ}\beta_I\wedge d\beta_J; K=
\begin{pmatrix} 2n+1&-1&0\\ -1&0&2\\ 0&2&0 \end{pmatrix}.
	\label{eqn:kmatrixstablaughlin} \end{align} One can construct a
 basis transformation $W$  of the form (c.f., Eq. \ref{eqn:equivKmatrix})
\begin{align} W&=    \begin{pmatrix} 2&2(1+2n)&1\\
1&n&0\\ -1&-(1+n)&0 \end{pmatrix};\quad W\in
GL(3,\mathbb{Z})\\ K&\rightarrow
WKW^T=\begin{pmatrix} 8n+4&0&0\\ 0&1&0\\ 0&0&-1
\end{pmatrix}
\label{eqn:Laughlinstabequivbasischange}
\end{align}
which reveals this theory to be equivalent to the state $K=8n+4$.
We see that this reproduces the action we found for Laughlin states with $K=8n+4$ modulo two
 fermion modes which can be gapped and lifted to higher energies. This also hints at how the fermionic TO was converted to a bosonic TO as the fermions can be trivially gapped. 

\section{Examples of Gauged Fermion Parity in FQHE States}\label{sec:examples}
In this Appendix we explicitly gauge the Fermion parity symmetry for a large class of fermionic FQHE states. 
\subsection{Hierarchy states} 
Abelian quantum hall states at a large number of observed filling fractions $\nu$ can
be described using hierarchy schemes \cite{Haldane83,Halperin84,Jainhierarchy}.
The general form of the $K$-matrix and charge vector of an electronic hierarchy state in
the Haldane-Halperin description takes the form \begin{align} K_{11}=1\,\text{
mod}\,\,2;&\quad K_{ii}=0\,\text{ mod}\,\,2; i>1\nonumber\\
\mathbf{t}^{T}&=(1,0,\cdots,0). \label{eqn:generalKhierarchy}
\end{align} The electron $\psi$ and $h/e$ flux $\mathcal{E}$ are given by the vectors
$K\mathbf{t}$ and $\bb{t}$ respectively (this choice ensures that the
$h/e$ flux quantum $\mathcal{E}$ has a charge of $\nu $ and statistical angle $\pi\nu$).  

The constraint, equation from our gauging procedure (Eq. 
\eqref{eqn:gaugeconstraint}) translates to \begin{align}
2b=t_I\alpha_I=\alpha_1.	\label{eqn:gaugeconstrainthierarchy}
\end{align} This implies that the bulk Lagrangian can be written as
\begin{align}
\mathcal{L}=\frac{1}{4\pi}\left({K_{\mbox{sc}}}\right)_{IJ}\alpha'_I\wedge
d\alpha'_J&\nonumber\\ \alpha'_I=\alpha_I,\,\,( I>1); \quad
2\alpha'_1=\alpha_1&\nonumber\\ K_{\mbox{sc}}=WKW^T;\quad
W=\delta_{ij}+\delta_{i1}\delta_{j1};\quad\alpha=W\alpha^{'}&\nonumber\\ 
W=\mbox{diag}(2,1,1,\cdots,1).\quad\quad\quad& \label{eqn:newKmatrix}
\end{align} Using Sylvester's law of inertia, we see that since $W$ is
invertible, the chirality of the system ($n_+-n_-)$, i.e., the difference
between the positive and negative eigenvalues of $K,$ is preserved under the
gauging process. Hence, the chiral central charge $c_-$ is preserved as expected.

The number of quasiparticles in the system is given by $\mbox{det}(K)$. Since,
$\mbox{det}(W)=2$ and $K_{\mbox{sc}}=WKW^T$, $\mbox{det}(K_{\mbox{sc}})=4\mbox{
det}(K)$. Thus, as expected from Eq. \eqref{eqn:degengauged}, and the fact that the resulting theory is Abelian, the number of
qps have increased by a factor of $4$. The new charge vector (of the now {\em local bosonic} state), is
\begin{align} \bb{t}_{\mbox{sc}}&=W\bb{t}\nonumber\\
\bb{t}_{\mbox{sc}}&=(2,0,\cdots,0)^T. \label{eqn:newchargevector}
\end{align}

Following Haldane \cite{Haldane95}, we rewrite the $K$ matrix of the
fermionic FQHE as $K= \begin{pmatrix} k_0&\bb{k}^T\\ \bb{k}&K_0 \end{pmatrix} $,
where $K$ is an $N\times N$ matrix, $\bb{k}$ is an $N-1$ dimensional
column vector, and $K_0$ is an $(N-1)\times(N-1)$ matrix. The charge vector
$\bb{t}$ takes the form $(t_0,0,\cdots,0)^T$. 
Hence for the gauged system we have \begin{align}
K_{\mbox{sc}}=WKW^T&= \begin{pmatrix} 4k_0&2\bb{k}^T\\ 2\bb{k}&K_0
\end{pmatrix};\nonumber\\
\bb{t}_{\mbox{sc}}=W\bb{t}&=(2t_0,0,\cdots,0)^T.
\label{eqn:fillingfracunchanged} \end{align} In the new basis $\alpha',$
the electron vector is $WK\bb{t},$ and $\m$ is $(1,0,\cdots,0)^T,$ i.e.,  $\m$
has unit charge under $b\equiv \alpha'_1.$ 

We can also check for the requisite properties of $\m$ and the electron:
\begin{align*}
\theta_{\bb{\psi,m}}&=2\pi
l_{\psi}^TK_{\mbox{sc}}^{-1}l_{\m}=2\pi
(WK\bb{t})^T(WKW^T)^{-1}\bb{t}\nonumber\\
&=2\pi\bb{t}^TW^{-1}\bb{t}\nonumber\\ &=2\pi \frac{1}{2}=\pi\\
q_{\m}&=(W\bb{t})^T(WKW^T)^{-1}\bb{t}=\bb{t}^TK^{-1}W^{-1}\bb{t}\nonumber\\
&=\frac{\bb{t}^TK^{-1}\bb{t}}{2}=\nu/2\\
\delta_\m&=\pi\bb{t}^{T}(WKW^T)^{-1}\bb{t}=\pi(W^{-1}\bb{t})^T
K^{-1}W^{-1}\bb{t}\\ =&\pi\frac{\bb{t}K^{-1}\bb{t}}{4}=\pi\nu/4
\end{align*} where
$l_{\bb{a}}$ is the qp vector for the qp $\bb{a}$. From this we see that  our physical expectations about the gauged theory, as outlined in Sec. 
\ref{sec:physicalconstraints}, are satisfied.  

For an explicit example, consider
the $\nu=2/5$  state described by the $K$ matrix and charge vector
\begin{align} K=\left( \begin{array}{cc} 3 & -1 \\ -1 & 2 \\
\end{array} \right);\bb{t}=\begin{pmatrix}{}1\\0\end{pmatrix}.
\label{eqn:twofifth} \end{align} Placing it in contact with an
s-wave superconductor and gauging the fermion parity we find an emergent state  described by
\begin{align} K_{\mbox{sc}}=\left( \begin{array}{cc} 12 & -2 \\
-2 & 2 \\ \end{array} \right); \qquad
t_{\mbox{sc}}=\begin{pmatrix}{}2\\0\end{pmatrix}.
\label{eqn:twofifthgauged} \end{align}  The
anyon fusion structure is better elucidated after
a basis transformation,(c.f. Eq. \eqref{eqn:equivKmatrix})
\begin{align} K_{\mbox{sc}}\rightarrow
W'K_{\mbox{sc}}W'^T=\left(
\begin{array}{cc} 10 & 0 \\ 0 & 2 \\
\end{array}
\right);&\,t_{\mbox{sc}}\rightarrow
W't_{\mbox{sc}}=\begin{pmatrix}{}2\\0\end{pmatrix};\nonumber\\
W'=\left( \begin{array}{cc} 1 &
1 \\ 0 & 1 \\
\end{array}\right).&
\label{eqn:anyonfusion2/5gauged}
\end{align} From this we see that the fusion structure is
clearly
$\mathbb{Z}_{10}\times\mathbb{Z}_{2}$.

The $\mathbb{Z}_2$ sector is neutral
and the charged $\mathbb{Z}_{10}$ sector is generated by the  $\m$
quasiparticle with $q_{\m}=e/5$. The original fermionic state had another $e/5$ qp to begin with  (denoted by $\bb{a}$) with
$\delta=3\pi/5$. The neutral sector is generated by the composite qp $\bb{a-m}$ which
has trivial statistics with $\m$ (i.e., the neutral qp made from $\bb{a}$ fused with the anti-particle of $\m$).

\subsection{Spin singlet state at $\nu=2/3$} 
The spin singlet state at filling $2/3$ is described by \begin{align} K=
\begin{pmatrix} 1&2\\ 2&1 \end{pmatrix};&\qquad t=(1,1).
\label{eqn:2/3Kmatrix} \end{align} The qp fusion group is
$\mathbb{Z}_3,$ and is generated by the quasiparticle $(1,0)^T$
with charge $e/3$ and statistical angle $\delta=2\pi/3$.  This
state was considered in the nice paper Ref. 
\onlinecite{clarke2014exotic} in the context of twist defects.

The application of Eq. \eqref{eqn:gaugeconstraint} leads to \begin{align}
2b=\alpha_1+\alpha_2.	\label{} \end{align} It is not immediately
clear how to simultaneously solve this constraint and obey the physical
conditions in Sec. \ref{sec:physicalconstraints}.  To do so we note that
$\bf{m}$ does not distinguish between spin up and spin down species as
far as braiding is concerned, as both carry a $\mathbb{Z}_2$ fermion
parity charge 1.  Now we implement a basis transformation
Eq. \eqref{eqn:equivKmatrix} which separates the system into a sector that is charged
under fermion parity and another sector which is neutral:
\begin{align} K\rightarrow WKW^T=\left( \begin{array}{cc} 1 & 1 \\ 1 &
-2 \\ \end{array} \right); &\bb{t}\rightarrow
W\bb{t}=(1,0)^T;\nonumber\\ \quad W=\left( \begin{array}{cc} 0
& 1 \\ 1 & -1 \\ \end{array} \right).
\label{eqn:equivKtwothird} \end{align} In this
representation, the new superconducting state can be
easily written down using our hierarchy state prescription:
\begin{align} K_{\mbox{sc}}&=\left(\begin{array}{cc} 4
& 2 \\ 2 & -2 \\ \end{array}\right);\quad
\bb{t_{\mbox{sc}}}=(2,0)^T.
\label{eqn:twothirdfpgauged} \end{align}
By a basis transformation, as in Eq. \eqref{eqn:anyonfusion2/5gauged}, it can be shown that its fusion 
is $\mathbb{Z}_6\times\mathbb{Z}_2$.
\subsection{$3/7$ Jain state} 

The Jain state at $\nu=3/7$ can be expressed using \begin{align} K=\left(
\begin{array}{ccc} 3 & 2 & 2 \\ 2 & 3 & 2 \\ 2 & 2 & 3 \\ \end{array}
\right);\,\, \bb{t}= \begin{pmatrix} 1\\1\\1	\end{pmatrix}.\nonumber\\
\label{eqn:threeseventhstate} \end{align} To consider
the emergent state in proximity contact with an s-wave superconductor after gauging fermion parity, we
perform a basis change into sectors which are fermion parity charged and
neutral \begin{align} K\rightarrow WKW^T
&=\left( \begin{array}{ccc} 3 & -1 & 0 \\ -1 &
2 & -1 \\ 0 & -1 & 2 \\ \end{array}
\right); \bb{t}\rightarrow W\bb{t}=
\begin{pmatrix} 1\\0\\0
\end{pmatrix}\nonumber\\ W=&\left(
\begin{array}{ccc} 0 & 0 & 1 \\ 0 & 1 &
-1 \\ 1 & -1 & 0 \\ \end{array}
\right).
\label{eqn:threeseventhbasischange}
\end{align} Now, following our prescription for gauging fermion parity 
for hierarchy states we can write down
the effective state easily
\begin{align} K_{\mbox{sc}} &=\left(
\begin{array}{ccc} 12 & -2 & 0
\\ -2 & 2 & -1 \\ 0 &
-1 & 2 \\ \end{array}
\right);\,
\bb{t_{\mbox{sc}}}=
\begin{pmatrix} 2\\0\\0
\end{pmatrix}.
\label{eqn:threeseventhfermionparitygauged}
\end{align}
\subsection{More Examples- the $ADE$ states}
In this section we will work with $K$ matrices of the following form:
\begin{align}
	K_{(r+1)\times(r+1)}&=
\begin{pmatrix}
2p\pm1 &  -1  & 0&\ldots& 0\\
-1   \\
0& &  \pm C_{r\times r}\\
\vdots  \\
0     
\end{pmatrix}\nonumber\\
t&=\begin{pmatrix}{}1,0,0,\cdots,0\end{pmatrix}^T
	\label{eqn:ADEseries}
\end{align}
Here $C$ is a $r\times r$ Cartan matrix corresponding to a rank $r$ Lie Algebra (see Appendix \ref{app:cartan}).
Since $C$ must be symmetric, the corresponding Lie Algebra must
be simply laced, hence we are restricted to the $A_n, D_n,$ or $E_n$ series.

The $A$ series is most relevant from the point of view of experimental
significance, as these states  directly relate to a number of stable quantum Hall states that 
have been observed at filling fractions
$\frac{n}{2np\pm 1}$. They have $u(1)\times su(n)_1$ symmetry in the $K$
matrix, where $su(n)$ corresponds to the $n-1$ dimensional Cartan matrix of
the $A$ series. These states have been studied thoroughly in the literature\cite{Readhierarchy,FrohlichZee,
FrohlichThiran94,FrohlichStuderThiran97,cappelli1995stable}. The $D$ series has been proposed in the context of even denominator quantum
Hall states\cite{Readhierarchy}, and the $E$ series is perhaps the least experimentally relevant currently.  

We should add some comment to avoid confusion with some recent literature. The K-matrices in Eq. \ref{eqn:ADEseries} have a fermionic block and a bosonic Lie algebra Cartan matrix. If we just considered the bosonic K-matrix by itself then we find another set of interesting bosonic FQHE states. 
The bosonic $D$ series has been in focus recently
in the context of 16 fold periodic classification\cite{Kitaev06} of topological
superconductors, and represents various versions of the toric code and
semion/anti-semion topological order. Additionally, the bosonic FQHE given by the $so(8)$
state relates to the surface states of some time reversal symmetric SPTS
\cite{BurnellChenFidkowskiVishwanath13,WangSenthil13,WangPotterSenthil13} and
non-Abelian anyonic symmetries.\cite{khan2014}
Finally, the bosonic $E_8$ state in particular is a bosonic
short-range entangled phase with no TO. \cite{LuVishwanathE8} Despite these interesting connections, our focus is purely on the fermionic hierarchy states in Eq.  \ref{eqn:ADEseries}.
%

\subsubsection{The Jain sequence at filling $\frac{n}{2np\pm 1}$; the $A$ series}
Quantum Hall states at filling $\frac{n}{2np\pm 1}$.
are described by an $n$-dimensional K-matrix (in the Jain construction
$n$ would correspond to the number of filled Landau
levels of the composite fermions/holes formed by attaching $2p$ flux quanta to
the bare electron). The anyon fusion group is $\mathbb{Z}_{2np\pm 1},$ and is generated by the $n$-dimensional quasiparticle vector $(0,\cdots,1)^T$ with charge
$\frac{e}{2np\pm 1}$.

Upon gauging the fermion parity symmetry, the resulting fusion group depends upon $n$.
If $n$ is odd, the fusion is $\mathbb{Z}_{4(2np\pm1)},$ and is generated by
the half quantum flux $\m$.
If $n$ is even, the fusion is given by $\mathbb{Z}_{4np\pm 2}\times\mathbb{Z}_2$.
The $\mathbb{Z}_{4np\pm 2}$ sector is charged and generated by $\psi-4p\m$ (with a
charge of $\frac{1}{(2np\pm 1)}$),
and the
neutral sector is generated by $(2np\pm 1)\m-\frac{n}{2}\psi$. 
However, note that these two
sectors are not completely decoupled in the sense that they
braid around each other with a phase of $-1$.

If $n=2$ mod 4, the sectors can be completely decoupled in the sense of
braiding too by using $\m$ to generate the charged $\mathbb{Z}_{4np\pm 2}$ sector.
This is what we saw for example in the $2/5$ state
which decoupled completely (the $K$-matrix had no off diagonal entries) into a $\mathbb{Z}_{10}$ and a $\mathbb{Z}_2$
sector (c.f., Eq. \eqref{eqn:anyonfusion2/5gauged}).

\subsubsection{$D$ series}
Let us now consider the $r+1$ dimensional $K$-matrix of the form Eq.
\ref{eqn:ADEseries} where now $C$ is the Cartan matrix corresponding to
$D_r\equiv so(2r)$.
The fermionic theory with the above $K$-matrix has filling fraction
$\nu=\frac{1}{2p}$. The anyonic fusion group is $\mathbb{Z}_{8p},$ and it is
generated by $(0,\cdots,0,1)^T$. Upon gauging fermion parity symmetry the
resulting theory has the anyonic fusion structure $\mathbb{Z}_{8p}\times
\mathcal{D}_{\pm r}$ (the $\pm$ sign refers to the element $K_{11}=2p\pm 1$
with the corresponding change in the sign of the Cartan matrix, as in Eq.
\ref{eqn:ADEseries}).
$\mathcal{D}_{\pm r}$ refers to the fusion structure of a purely bosonic theory with
Cartan matrix $\pm D_r$.\cite{khan2014,mathieu1997conformal}. The sector
$\mathcal{D}_r$ is
completely neutral and has fusion $\mathbb{Z}_4$ or $\mathbb{Z}_2\times\mathbb{Z}_2$
depending on whether $r$ is odd/even. The sector $D_{-r}$ can be understood as
having time reversed braiding with respect to $D_{r}$.

If $r$ is odd $\mathcal{D}_r$  is
generated by $(\mp 1,\cdots,1)^T$ which is neutral and has $h=\frac{\pm r}{8}$.
If $r$ is even there are two neutral generators $(\mp 1,\cdots,1)^T$(with
$h=\pm \frac{r}{8}$) and 
$4p\bb{\m-\psi}$ (with fermionic statistics). 

The charged sector $\mathbb{Z}_{8p}$ is a $U(1)_{4p}$ theory. It can be
expressed by a single component $K$-matrix with $K=8p,$ and  $\mathbf{t}=2$. It decouples
completely from the $\mathcal{D}_r$ sector, and is generated by the half quantum
flux $\m$. 

Interestingly, in the $D$ series there is spin and charge separation, i.e., the topological state splits up into two parts, a bosonic $U(1)_{4p}$ sector which carries all the electric charge, and a neutral $\mathcal{D}_r$ sector that contains a fermionic particle 
$f$ which supports the statistics/spin of the electron. $f$ can be represented by a vector $(\mp 2,1,0,\cdots,0),$ and the electron field decomposes into $\psi=4p{\bf m} + f$, where $4p\m$ is the bosonic charged part.
\subsubsection{$E$ series}
Here the dimension of the full $K$-matrix is $7, 8$, or $9$ depending on whether the Cartan
matrix in Eq. \eqref{eqn:ADEseries} is $E_6, E_7,$ or $E_8$.

For $E_6$, before gauging fermion parity the fusion group is
$\mathbb{Z}_{6p\mp 1}$, the filling fraction is $\frac{3}{6p\mp 1},$ and the qps are generated
by the $h/e$ flux.
Upon gauging fermion parity, the fusion group is $\mathbb{Z}_{4(6p\mp 1)}$ and is generated
by the half quantum flux $\bb{m}$.

In the case of $E_7$, the filling fraction is $\nu=\frac{1}{2p\mp 1}$, and the fusion group
is $\mathbb{Z}_2\times\mathbb{Z}_{2p\mp 1}$. The 
$\mathbb{Z}_{2p\mp 1}$ sector is charged and is generated by the $h/e$ flux, the $\mathbb{Z}_2$ sector is neutral
and is generated by the the quasiparticle vector $(0,1,0,0,0,0,0,-1)^T$. The generator of
the neutral sector has $h=\pm\frac{3}{4}$ as well. The two
sectors are completely decoupled from each other. After gauging fermion parity the
fusion group becomes $\mathbb{Z}_{4(2p\mp 1)}\times \mathbb{Z}_2$. The
$\mathbb{Z}_2$ sector is still neutral and stays unaffected.  Only the charge
sector gets quadrupled and is now generated by $\m$.

For $E_8$ the filling fraction is $\frac{1}{2p\mp 1}$. The fusion group is
$\mathbb{Z}_{2p\mp 1}$ and is generated by $h/e$. After gauging fermion parity, the theory is now $\mathbb{Z}_{4(2p\mp 1)}$ and is generated
by $\m$.

\section{Making the fermion parity flip more relevant}\label{app:relevance}
This Appendix focuses on interactions at a non-chiral quasi-one dimensional interface appropriate
to a trench (see Figures \ref{fig:SCTOlayer} \& \ref{fig:edgecouplings} ) in a
TO state-s-wave SC heterostructure device. 
We have seen that a MBS or a parafermion is localized at a twist
defect depending on whether fermion parity flip AS $\m\rightarrow \m\times\psi$ is more 
relevant than the charge conjugation symmetries $\m\rightarrow \m^{-1}$ and
$\m\rightarrow \m^5$
for the case of the Laughlin 1/3
state. 
The main objective of this section is to show that the nature of the NAZM
depends on the interactions near the cut. In particular the scaling dimension
of the backscattering/gapping term in Eq. \ref{interfaceK} depends on the velocity
matrix $V$ which encodes the forward-scattering between the edges.
In particular we show the  {\emph
{existence}} of a velocity matrix $V$ which favors MBS over parafermions.
Finding precise specifications for experimentally accessible regions of forward scattering
interactions that meet this criterion is a much more difficult task and we
leave that for future investigations.

We first set up conventions and
outline the procedure to calculate the scaling dimensions of a backscattering
term
in a general TO state using
the method in Refs. \onlinecite{MooreWenedgerelevance1,MooreWenedgerelevance2}. 
We begin with the Lagrangian density at the edge, from Sec. \ref{sec:AS}:
\begin{align}
	\mathcal{L}&=\mathcal{L}_{\text{edge}}+\mathcal{L}_{\text{gap}}
	\nonumber\\
	\mathcal{L}_{\text{edge}}&=\frac{1}{4\pi}\left[{\left(K\oplus
	-K\right)}^{\alpha\beta}_{IJ}\partial_x\phi^{I}_{\alpha}\partial_t\phi^{J}_\beta
	-V^{\alpha\beta}_{IJ}\partial_x\phi^{I}_{\alpha}\partial_x\phi^{J}_\beta\right]\nonumber\\
	\mathcal{L}_{\text{gap}}&=-g_I^W\cos\left[K_{IJ}\left({\pmb
	\phi}_b+W^T{\pmb \phi}_t\right)_J\right]
\label{eqn:lagdensityphi}
\end{align} 
where, $\alpha,\beta$ run over $t,b$ denoting the edge modes at the top and
bottom sides of the interface respectively and $I,J=1,\cdots,N$ (note that there are a total of $2N$
modes at the interface, $N$ from the top/bottom). The bold symbols denote the
column vector $\pmb{\phi}$ with elements $\phi^I$.

We now implement a basis transformation ${\pmb \phi}=M_1{\pmb X}$, so that 
 the $K$ matrix at the edge is transformed to the pseudo-identity $I_N\oplus-I_N$. 
In this basis, the first $N$ modes with signature $+$ are right moving and the
last $N$ with signature $-$ are left moving. This is appropriate since we will eventually specialize to a Laughlin state which is fully chiral on a single edge. We will decompose the $2N$
dimensional vector $\pmb{X}=
\begin{pmatrix}
	\pmb{X_R}\\
	\pmb{X_L}
\end{pmatrix}$.
The Lagrangian density takes
the form
\begin{align}
	\mathcal{L}&=\mathcal{L}_{\text{edge}}+\mathcal{L}_{\text{gap}}
	\nonumber\\
	\mathcal{L}_{\text{edge}}=\frac{1}{4\pi}[{\left(I_N\oplus
	-I_N\right)}_{IJ}&\partial_x X^{I}\partial_t X^{J}_\beta
	-V^X_{IJ}\partial_x X^{I}\partial_x X^{J}]\nonumber\\
	\mathcal{L}_{\text{gap}}&=-g_I^W\cos\left(R_I^T\mathbf{X_R}+L_I^T\mathbf{X_L}\right)\nonumber\\
	M_1^TVM_1=V^X; & \left(R^T\quad L^T\right)= \left(W^{-1}K \quad
	K\right) M_1
\label{eqn:lagdensityX}
\end{align}
\noindent where we have added possible gapping terms, and $R_I/L_I$ corresponds
to the $I$-th column of the matrix $R/L$. 
Note that $R,L,W,K$ are $N\times N$ matrices while $M_1$ and $V$ are $2N\times 2N$.
Haldane's null vector criterion \cite{Haldane95} for gapping terms implies that 
\begin{align}
	R^{T}R=L^{T}L.	
\label{eqn:haldanenullvector}
\end{align}
Now, let us assume that the matrix $V^{X}$ can be diagonalized by another basis
change without affecting the pseudoidentity K-matrix $I_N\oplus -I_{N}$ using the transformation $X=O\tilde{X}$.
In this new basis, the forward scattering is encoded in $\tilde{V}=O^TV^{X}O$. The
gapping term $\cos[\mathbf{a}^T\pmb{X}]$ is now $\cos[\mathbf{\tilde
a}^T\tilde{\pmb{X}}]$ where $\tilde{\mathbf{a}}=O^T\mathbf{a}$.
The scaling dimension of the operator $e^{i\tilde{\mathbf{a}^T}\pmb X}$ is
\begin{align}
\langle e^{i\tilde{\mathbf{a}^T}\pmb X}e^{-i\tilde{\mathbf{a}^T}\pmb
X}\rangle&=\langle e^{i\tilde{\mathbf{a}_L^T}\pmb X_L}e^{-i\tilde{\mathbf{a}_L^T}\pmb
X_L}\rangle\langle e^{i\tilde{\mathbf{a}_R^T}\pmb X_R}e^{-i\tilde{\mathbf{a}_R^T}\pmb
X_R}\rangle\nonumber\\
&=\prod_{j=1}^{3}\frac{1}{\left(\tilde{v}_{J,R}t-x\right)^{\tilde{a}_{R,J}^2}}
\prod_{j=1}^{3}\frac{1}{\left(\tilde{v}_{J,L}t+x\right)^{\tilde{a}_{L,J}^2}}
\label{eqn:scalingdim1}
\end{align}
where, $\tilde{v}_{J,R/L}$ denote the velocities of the right/left moving modes.
Thus, the scaling dimension is
\begin{align}
	\Delta({\pmb a})&=\frac{1}{2}\tilde{\pmb a}^T\pmb{\tilde
	a}=\frac{1}{2}{\pmb a}^TOO^T\pmb{a}.	
\label{eqn:scalindim2}
\end{align}

Now, we write $O=BR$ using the result from Refs. \onlinecite{MooreWenedgerelevance1,MooreWenedgerelevance2} that any
matrix $O\in SO(N,N)$ can be written as the product of a symmetric positive boost $B$ and a
rotation $R$, such that $B,R$ are both in $SO(N,N)$. Hence, $B^2=OO^T$.
Thus, the scaling dimension becomes
i\begin{align}
	\Delta({\pmb a})&=\frac{1}{2}{\pmb a}^TOO^T\pmb{a}=\frac{1}{2}{\pmb
	a}^TB^2\pmb{a},		
\label{eqn:scalindim}
\end{align}
and we see the scaling dimension depends only on the boost. The physical
picture is that the scaling dimensions are independent of the velocities of the
eigenmodes and the interactions between co-propagating modes. They are
dependent only on the interactions between counter-propagating modes.
We now parametrize the boost $B$ as 
i\begin{align}
B=\exp
\begin{pmatrix}
0&b/2\\
b^T/2&0
\end{pmatrix}
\label{eqn:boostparam}
\end{align}
where, $b$ is an $N\times N$ dimensional real-valued matrix.
Note now, that $B$ has been parametrized by the $N^2$ parameters of the matrix
$b$. This is just an $N+N$ dimensional generalization of Lorentz boosts in $1+1$-d.
At the interface with $2N$ edge modes, there are $N$ gapping terms. 
Now, given a gapping term $W,$ and hence corresponding matrices $R$ and $L,$ and
using
Eq. \ref{eqn:scalindim}, we see that their scaling dimensions can be
obtained from the diagonal elements of the matrix product
\begin{align}i
	\delta_I=\frac{1}{2}\left[\begin{pmatrix}R^T L^T\end{pmatrix}
		B^2 \begin{pmatrix}R\\ L\end{pmatrix}\right]_{II}.	
\label{eqn:scalingmatrixform}
\end{align}

 Before going further, we give an example of a velocity matrix
$V^{X}$ which is diagonalized by elements $O$ in $SO(N,N)$. Let us consider,
the positive definite velocity matrix 
\begin{align*}
	V^{X}&=\exp
\begin{pmatrix}
0&-b\\
-b^T&0
\end{pmatrix};
\end{align*}
It is diagonalized by the transformation $O^TV^{X}O$ using $O=B$, with $B$ from
Eq. \eqref{eqn:boostparam}.  

Now, we will make a particular set of gapping terms (characterized by the matrices $R$
and $L$) relevant. We  
borrow the ansatz of Ref. \onlinecite{CanoMulliganHughes} and take $b=\alpha RL^{-1}$.
In the following equation we make this choice for $b$, and denote
the corresponding value of $B^2$ as $B^2(b=\alpha RL^{-1})$.
Using Eq. \ref{eqn:haldanenullvector},
\begin{align*}
	B^2=\sum_{n=0}^{\infty}\frac{\alpha^{2n}}{(2n) 
	!}I+\frac{\alpha^{2n}}{(2n+1)!}	
	\begin{pmatrix}
	0&b\\
	b^T&0
	\end{pmatrix}.	
\end{align*}
Finally, with the help of Eq.  \ref{eqn:scalingmatrixform} we get the scaling
dimensions of the gapping terms
$\delta_I(b=\alpha RL^{-1})=\exp(\alpha )\left[L^TL\right]_{II}$.
Thus, by choosing $\alpha$ large and negative we can make all three gapping
terms simultaneously relevant. In 1+1-d this happens when $\Delta_I$
is less than 2 for each of the gapping terms.

Now, we specialize to the Laughlin $\nu=1/3$ state along with a proximity-coupled s-wave superconductor and a trench interface. We have seen in Sec. 
\ref{sec:AS} that realizing the fermion parity flip symmetry in the K-matrix formalism  requires extending
the usual $K$ matrix to consider $8n+4\oplus\sigma_x$ instead. Since the $K$-matrix is
$3\times 3$, there should be $3$ gapping terms at the interface. Since we only want to show a proof of principle, we have freedom in our choice of forward scattering interactions. Let us choose
$B$ of the form Eq. \ref{eqn:boostparam}  with $b=-6R^{-1}L$, i.e. $\alpha=-6$,
where $R$ and $L$ are determined using Eq. \ref{eqn:lagdensityX} using
$W_{\text{fpf}}$ appropriate to fermion parity flip from Eq. 
\ref{eqn:1/3rdstablequiv}. With this form of the velocity matrix, we find that
the fermion parity flip AS $\m\rightarrow \m^7$ is relevant, as designed, while the charge conjugation symmetries 
$\m\rightarrow \m^5$ and $\m\rightarrow \m^{-1}$ are  irrelevant.  The numerical values of the scaling
dimensions of the 3 gapping terms for the fermion parity flip are
$\{1.63598,0.151204,0.0322238\}$, and are all less than 2, whereas those for the charge
conjugation symmetry are \{3025.73, 302.572, 302.572\}. For the composite
symmetry $\m\rightarrow \m^5$ it is \{4842.75,554.862,252.174\}, making the latter two very
irrelevant. Hence, we have found an interaction which would prefer to have MBS twist defects instead of parafermion twist defects.

\section{Cartan Matrices for the $ADE$ series}\label{app:cartan}
Here we provide the Cartan matrices of the $ADE$ series used in the construction of
quantum Hall states outlined in the main text.

The Lie algebras $A_r=su(r+1)$, for $r\geq2$, and $D_r=so(2r)$, for $r\geq4$, can be used to form infinite series of Abelian states, each with a Cartan matrix of rank $r$: 
\begin{align}
	\left(K_{A_r}\right)_{IJ}&=2\delta_{IJ}-(\delta_{I,J+1}+\delta_{I,J-1})\label{eqn:CartanA}\\
\left(K_{D_r}\right)_{IJ}&=2\delta_{IJ}-(\delta_{I,J+1}+\delta_{I,J-1})+\nonumber\\
&\left(\delta_{I,r}\delta_{J,r-1}+\delta_{I,r-1}\delta_{J,r}-\delta_{I,r}\delta_{J,r-2}-\delta_{I,r-2}\delta_{J,r}\right).\label{eqn:CartanD}
\end{align}

There are three exceptional simply-laced Lie algebra $K_{E_{r=6,7,8}}$ with Cartan matrices \begin{align}K_{E_6}=\left(
\begin{array}{cccccc}
 2 & -1 & 0 & 0 & 0 & 0 \\
 -1 & 2 & -1 & 0 & 0 & 0 \\
 0 & -1 & 2 & -1 & 0 & -1 \\
 0 & 0 & -1 & 2 & -1 & 0 \\
 0 & 0 & 0 & -1 & 2 & 0 \\
 0 & 0 & -1 & 0 & 0 & 2 \\
\end{array}
\right),
\label{eqn:CartanE6}
\end{align}

\begin{align} K_{E_7}=\left(
\begin{array}{ccccccc}
 2 & -1 & 0 & 0 & 0 & 0 & 0 \\
 -1 & 2 & -1 & 0 & 0 & 0 & 0 \\
 0 & -1 & 2 & -1 & 0 & 0 & -1 \\
 0 & 0 & -1 & 2 & -1 & 0 & 0 \\
 0 & 0 & 0 & -1 & 2 & -1 & 0 \\
 0 & 0 & 0 & 0 & -1 & 2 & 0 \\
 0 & 0 & -1 & 0 & 0 & 0 & 2
\end{array}
\right
),
\label{eqn:CartanE7}
\end{align}

\begin{align}
K_{E_8}=
\left(
\begin{array}{cccccccc}
 2 & -1 & 0 & 0 & 0 & 0 & 0 & 0 \\
 -1 & 2 & -1 & 0 & 0 & 0 & 0 & 0 \\
 0 & -1 & 2 & -1 & 0 & 0 & 0 & 0 \\
 0 & 0 & -1 & 2 & -1 & 0 & 0 & 0 \\
 0 & 0 & 0 & -1 & 2 & -1 & 0 & -1 \\
 0 & 0 & 0 & 0 & -1 & 2 & -1 & 0 \\
 0 & 0 & 0 & 0 & 0 & -1 & 2 & 0 \\
 0 & 0 & 0 & 0 & -1 & 0 & 0 & 2 \\
\end{array}
\right).
\label{eqn:CartanE8}
\end{align}

%

\end{document}